\documentclass[aps,preprint]{revtex4-1}
\usepackage[bookmarksnumbered=true,
	   	 colorlinks=true,
            linkcolor=blue,
            citecolor=red,
            urlcolor=blue
            ]{hyperref}
\usepackage[left=2.5cm, right=2.5cm, top=2.5cm, bottom=2.2cm]{geometry}
\usepackage{natbib} 
\usepackage{bm}
\usepackage{amsfonts,amssymb} 
\usepackage{amsmath} 
\usepackage{commath} 
\usepackage{xcolor}
\usepackage{tikz}
\usepackage{overpic}
\usepackage[titletoc,title]{appendix}
\numberwithin{equation}{section} 
\usepackage[noabbrev]{cleveref} 
\crefname{appsec}{appendix}{appendices} 
\begin{document}
\title{Trapping and displacement of liquid collars and plugs in rough-walled tubes}

\author{Feng Xu}
\email{f.xu.2@bham.ac.uk}
\affiliation{School of Mathematics, University of Birmingham, Edgbaston, Birmingham B15 2TT, UK}
\author{Oliver E. Jensen}
\email{oliver.jensen@manchester.ac.uk}
\affiliation{School of Mathematics, University of Manchester, Oxford Road, Manchester M13 9PL, UK}
\date{\today}

\begin{abstract}
A liquid film wetting the interior of a long circular cylinder redistributes under the action of surface tension to form annular collars or occlusive plugs.  These equilibrium structures are invariant under axial translation within a perfectly smooth uniform tube and therefore can be displaced axially by very weak external forcing.  We consider how this degeneracy is disrupted when the tube wall is rough, and determine threshold conditions under which collars or plugs resist displacement under forcing.  Wall roughness is modelled as a non-axisymmetric Gaussian random field of prescribed correlation length and small variance, mimicking some of the geometric irregularities inherent in applications such as lung airways.  The thin film coating this surface is modelled using lubrication theory.   When the roughness is weak, we show how the locations of equilibrium collars and plugs can be identified in terms of the azimuthally-averaged tube radius; we derive conditions specifying equilibrium collar locations under an externally imposed shear flow, and plug locations under an imposed pressure gradient.  We use these results to determine the probability of external forcing being sufficient to displace a collar or plug from a rough-walled tube, when the tube roughness is defined only in statistical terms.   
\end{abstract}


\maketitle

\section{Introduction}
\label{sec:introduction}

Surface-tension driven flows in liquid-lined lung airways are of physiological importance in driving airway closure (through the Rayleigh instability, causing redistribution of airway liquid into occlusive liquid plugs, and through compressive loading of capillary forces on the flexible airway wall) and subsequent airway opening (by displacement of liquid plugs and airway inflation).  Significant insight into the physical processes underlying mechanisms of airway closure and reopening have come from studies of idealised model problems, many of which have wider relevance to two-phase flow in porous media and microfluidics.  In particular, models based on lubrication theory of the initial Rayleigh instability in a uniform liquid-lined tube, describing the formation of annular collars \citep{1983-Hammond-p363} and liquid bridges (or plugs) \citep{1988-Gauglitz-p1457}, have been extended to account for numerous features of the complex \textit{in vivo} airway environment in health and disease, including factors such as wall elasticity and airway collapse, surfactants, imposed shear due to airflow, gravity, airway centreline curvature and non-Newtonian rheology (reviewed in \cite{2001-Grotberg-p421, 2004-Grotberg-p121, 2005-Bertram-p1681, 2008-Heil-p214, 2011-Heil-p141, Grotberg2011resp, 2014-Levy-p107}).

The present study addresses a curious aspect of the original problem studied by \citet{1983-Hammond-p363}.  A thin, initially uniform liquid film coating the interior of a long circularly-cylindrical tube, subject to no-flux conditions applied at either end of the tube, redistributes under the action of surface tension into a set of annular collars, connected to each other by a slowly draining film that remains continuous in the absence of evaporation and destabilising disjoining pressure.  While the collar shape is well approximated as a surface of uniform curvature (an unduloid that meets the tube wall with zero contact angle \citep{1972-Everett-p125}), the collar location is not so readily determined.  In practice, collars will either stabilize at either end of the tube (with centres coincident with no-flux boundaries), or they may migrate towards a boundary, even reversing direction over very long time-scales under the influence of very small differences in the draining flows near each edge of the collar  \citep{2006-Lister-p311}.  This sensitivity reflects a degeneracy in the underlying model, whereby an equilibrium wetting collar (or, if sufficient fluid is available, an occlusive liquid plug) can in principle be located anywhere along a sufficiently long uniform circular cylinder.   
 
In reality, lung airways (and other tubes arising in natural environments) are not perfectly uniform cylinders.  Major geometrical imperfections (such as centreline curvature) have already been shown to disrupt collars sufficiently to form axial rivulets \citep{1997-Jensen-p373, 2012-Hazel-p213}.  Here we address a more subtle distortion, imagining that the tube has small random perturbations to its shape (in the case of an airway, arising for example from protruding epithelial cells, inhaled debris {{or mucosal buckling}}), which are described by a Gaussian random field of prescribed variance and correlation length.  We seek the conditions under which such perturbations are sufficient to stabilize collars at discrete locations along the tube.  Small shape perturbations represent a singular limit of Hammond's problem, regularizing the degeneracy associated with axial collar translation.   
 
We characterise the wall shape in statistical terms (representing the uncertainties and natural variability that are inherent to physiological systems) and correspondingly express outcomes in probabilistic terms.  We build on prior deterministic studies of the influence of axially periodic axisymmetric corrugations along the tube on collar formation, addressing linear \citep{2002-Wei-p113} and weakly nonlinear \citep{2002-Wei-p149} stability as well as the fully nonlinear dynamics leading to tube occlusion \citep{2010-Beresnev-p12105, 2016-Wang-p500}, which typically show alignment of near-equilibrium structures with tube constrictions.  Non-axisymmetric perturbations to the exterior of a liquid-lined cylinder can also influence film distributions \cite{li2017}.  Our study complements the extensive literature on liquid plugs in lung airways, which focuses primarily on plug displacement under forcing, with studies focusing on the role of inertia \citep{fujioka2004}, wall flexibility \citep{howell2000, zheng2009}, surfactant \citep{waters2002, fujioka2005}, dynamical effects including plug rupture \citep{fujioka2008, ubal2008, hu2015, magniez2016}, gravity \citep{suresh2005, zheng2007}, interactions with bifurcations \citep{zheng2006,vaughan2016} and yielding behaviour \citep{zamankhan2012a,jalaal2016}. 

This study extends this prior work by addressing the role of random geometric imperfections on collar and plug dynamics.  For tubes having weak non-axisymmetric roughness, we show how equilibrium collar and plug locations are defined in terms of the azimuthally-averaged wall shape, motivating the study of axisymmetric tubes with axially non-uniform shapes.  For such tubes, having randomly distributed rather than periodic constrictions, we derive algebraic conditions for the existence of stable capillary equilibria.  We use two interpretations of stability: one involving perturbations driven by surface tension effects alone; the other involving an imposed external perturbation (here we consider shear from a core flow displacing collars, or an external pressure gradient displacing occlusive liquid plugs).  For applications in which the tube shape is defined only at a statistical level, we then determine the probability that, over many realisations, a given external forcing is sufficient to displace an isolated collar or plug.  We first present results for collars (Secs.~\ref{sec:problem}--\ref{sec:critshear}) and briefly discuss extensions to liquid plugs in Sec.~\ref{sec:plugs} below, giving displacement probabilities in an explicit analytic form.

\section{Model for slender liquid collars}
\label{sec:problem}

\begin{figure}
\begin{center}
\begin{overpic}[width=0.9\linewidth]{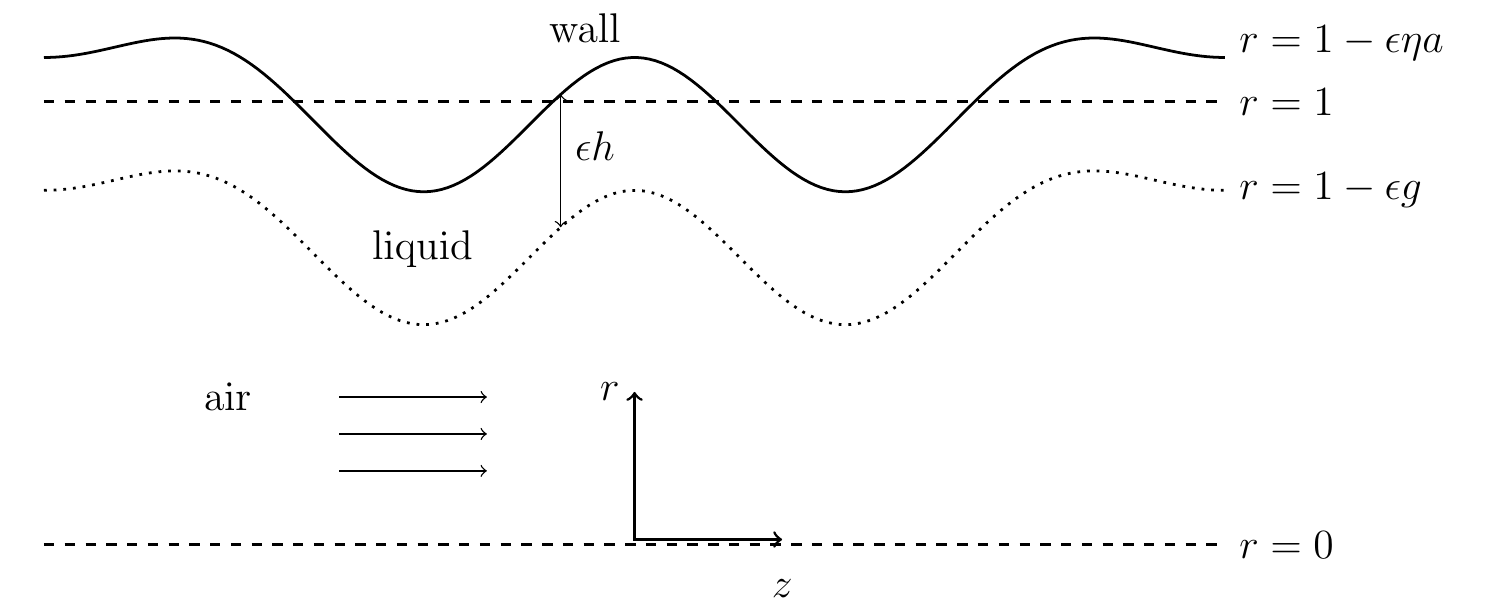}
\end{overpic}
\end{center}
\caption{A sketch of the problem domain showing axial $a_0 z$ and radial coordinates $a_0 r$ at fixed azimuthal angle $\theta$, showing wall location $r=1-\epsilon \eta a(\theta,z)$ (solid), film thickness $\epsilon h(\theta,z,t)$ and interface location $r=1-\epsilon g(\theta,z)$ (dotted).  The dashed line represents the tube centreline.  {The wall shape is arbitrary in this illustration.}}
\label{fig:1}
\end{figure}

We consider a rigid hollow cylinder with a rough interior wall coated by a thin layer of fully-wetting Newtonian liquid of viscosity $\mu$ and surface tension $\sigma$ (Figure~\ref{fig:1}). We assume the wall roughness amplitude and the liquid-layer thickness are of comparable magnitude and that both are much smaller than the cylinder's mean internal radius $a_0$.  We express the liquid-layer thickness as $\epsilon a_0 h(\theta, z,t)$ and the internal radius of the cylinder as $a_0(1-\epsilon \eta a(\theta,z))$, where $a$ is a realisation of a Gaussian random field having zero mean and $O(1)$ exponential covariance.  Here $\epsilon \ll 1$ is the ratio of the spatially-averaged liquid-layer thickness to $a_0$, $\eta$ is the ratio of the roughness amplitude relative to the liquid-layer thickness, $\theta \in[0,2\pi)$ measures the azimuthal angle around the tube, $a_0 z$ measures axial distance and time $t$ is scaled on $a_0 \mu/\epsilon^3 \sigma$.  The tube has length $a_0 L$.  The liquid layer is subject to an imposed axial shear stress $\epsilon^2 \sigma \tau /a_0$ (due to flow of gas in the core of the tube) and satisfies no-slip and no-penetration conditions at the cylinder wall. We neglect gravity and inertia and assume the cylinder's length $a_0L$ significantly exceeds its radius.

In the leading-order lubrication approximation, the dimensionless liquid-layer thickness $h(\theta, z, t)$ satisfies \cite{1997-Jensen-p373, 2006-Lister-p311}
\begin{align}\label{EvolutionEq}
h_t -\left(\tfrac{1}{3}h^3 p_\theta\right)_\theta + \left(\tfrac{1}{2} \tau h^2 - \tfrac{1}{3}h^3 p_z\right)_z = 0, \quad p = -(g + g_{\theta\theta}+g_{zz}), \quad g = h+ \eta a(\theta, z),
\end{align}
where subscripts denote derivatives.  The film adjusts under the action of the imposed shear stress $\tau$ and pressure gradients arising from the non-uniform curvature of the gas-liquid interface (Figure~\ref{fig:1}).  The function $g(\theta,z,t)$ measures the distance of this interface from the mean wall location; in the expression for the linearized mean curvature in (\ref{EvolutionEq}b), the term $g$ represents the azimuthal curvature that drives collar formation.  We impose periodic boundary conditions for $h(\theta, z, t)$ around the boundary of the domain $[0, 2\pi] \times [0,L]$, ensuring conservation of fluid volume \begin{equation}
V_0=\int_0^{2\pi} \int_0^L h\, \mathrm{d}z\mathrm{d}\theta.
\end{equation}
(Axial periodicity is imposed for computational convenience but later we will relax this condition.)  We impose $h=1$ at $t=0$, so that $V_0=2\pi L$.  Since disjoining pressure effects are neglected, the film remains continuous with $h>0$ everywhere, although it can become very thin over much of the domain at large times.

\begin{figure}
\begin{center}
\begin{overpic}[width=0.9\linewidth]{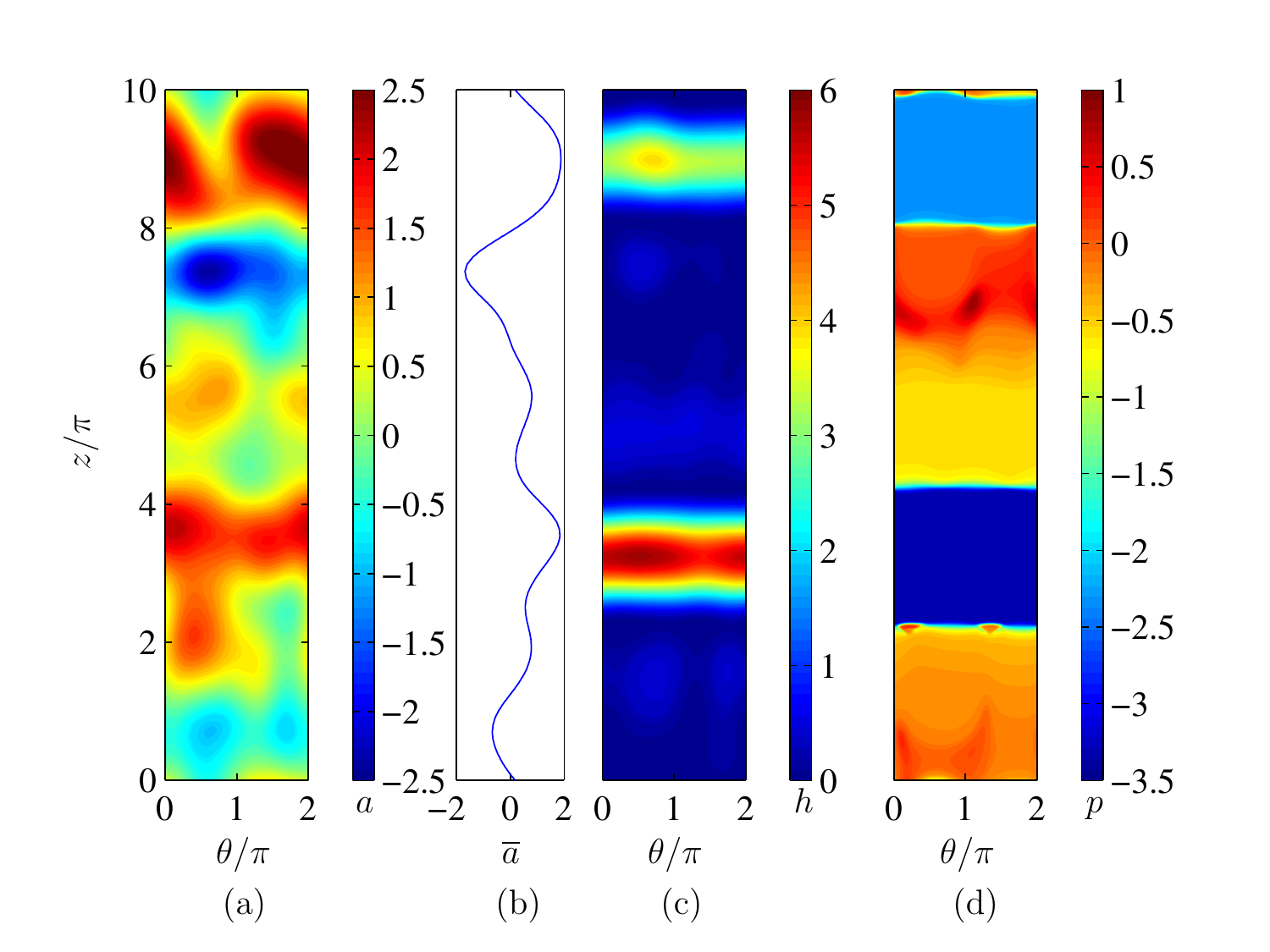} 
\end{overpic}
\label{2DCollar}
\end{center}
\caption{A simulation from initial condition $h=1$ of the spatially two-dimensional evolution equation (\ref{EvolutionEq}) for $\tau=0$, $\eta=0.5$, $L=10\pi$, $r_1=r_2=2$ showing the formation of equilibrium collars: (a) a realisation of the doubly-periodic wall roughness $a(\theta,z)$; (b) its azimuthal average $\overline{a}(z)$; (c) the near-equilibrium film thickness $h(\theta,z,1000)$; (d) the pressure $p(\theta,z,1000)$.}
\label{fig:0}
\end{figure}

In order to be compatible with the periodic boundary conditions, we consider the wall roughness field $a(\theta, z; \omega)$ to be a doubly periodic stationary Gaussian random field with zero mean and covariance function 
\begin{align}\label{covariance}
k(\theta,\theta';z,z') & = \textrm{exp}\left[-\frac{1}{2}\left[\left(\frac{\sin((\theta - \theta')/2)}{r_1/2}\right)^2+\left(\frac{\sin(\pi(z - z')/L)}{\pi r_2/L}\right)^2\right]\right],
\end{align}
where $r_1$ and $r_2$ are $O(1)$ dimensionless correlation lengths associated with the $\theta$ and $z$ directions, comparable in magnitude to the cylinder radius {{($\min(r_1,r_2)\gg \epsilon\eta$ ensures that the assumptions of lubrication theory are not violated)}}.   The $\sin^2$ form in each direction arises from sampling a planar 2D field with exponential covariance around a circle \citep{rasmussen2006}. Here we allow $r_1$ and $r_2$ to be varied independently; our approach below readily accommodates other choices of $k$, such as  
\begin{align}\label{covariancecyl}
k_{\mathrm{iso}}(\theta,\theta';z,z') = \textrm{exp}\left[-\frac{1}{2}\left[\left(\frac{\sin((\theta - \theta')/2)}{r/2}\right)^2+\left(\frac{z - z'}{ r }\right)^2\right]\right],
\end{align}
which represents the restriction of a 3D isotropic Gaussian field with correlation length $r$ to a long cylinder of unit radius.   $\omega$ identifies $a(\theta,z;\omega)$ as a sample drawn from a probability distribution and it implicitly labels realisations of the model, although we largely suppress it in what follows.  We use a Karhunen--Lo\'{e}ve decomposition to generate samples of $a$ numerically \citep{Lord2014}.  The associated film distribution is obtained by solving (\ref{EvolutionEq}) using a finite-element method, implemented in COMSOL Multiphysics, with care taken to resolve fine-scale structures that emerge at large times.  A realisation of $a(\theta,z)$, and its azimuthal average $\overline{a}(z)$, are shown in Figure~{\ref{fig:0}(a,b)}, where the azimuthal averaging operator is defined as
\begin{align}
\overline{f} = \frac{1}{2\pi}\int_0^{2\pi}f d \theta.
\label{eq:axiav}
\end{align}

The interfacial area (equivalent to a capillary surface energy) associated with (\ref{EvolutionEq}) can be expressed (to the appropriate level of accuracy) as 
\begin{equation}
\mathcal{A}(t)=\int_0^{2\pi} \int_0^L \left[\tfrac{1}{2} (h_z^2+h_\theta^2-h^2) - \eta(ah-a_z h_z - a_\theta h_\theta)\right] 
\, \mathrm{d}z\mathrm{d}\theta.
\label{eq:intenergy}
\end{equation}
Imposing periodic boundary conditions, integration by parts demonstrates that
\begin{equation}
\mathcal{A}_t=\int_0^{2\pi}\int_0^L h_t p\,\mathrm{d}z\mathrm{d}\theta=\int_0^{2\pi}\int_0^L \left[ \tfrac{1}{2} \tau h^2 p_z-\tfrac{1}{3}h^3 p_\theta^2 - \tfrac{1}{3} h^3 p_z^2 \right] \,\mathrm{d}z\mathrm{d}\theta.
\label{eq:at}
\end{equation}
Imposed shear injects energy into the system, such that $\mathcal{A}$ can be expected to evolve to a statistically steady state at large times.  However in the absence of shear, (\ref{eq:at}) shows that $\mathcal{A}_t\leq 0$, so that $\mathcal{A}$ continually diminishes as the film evolves, although typically the system approaches a metastable state that does not represent the global energy minimum of the system.  In a perfectly cylindrical tube ($\eta=0$), the global minimum has the entire fluid volume confined within a single equilibrium collar of the form
\begin{equation}\label{collarsol0}
h\approx h_0\equiv  ({V_0}/{4\pi^2})(1+\cos(z-z^\star)) \qquad (\vert z-z^\star\vert<\pi) 
\end{equation}
where the collar may lie anywhere in the domain; elsewhere $h$ is vanishingly small but non-zero.  The collar described by (\ref{collarsol0}) has zero contact angle at each of its effective contact lines ($h_0 = h_{0z} = 0$ at $z =z^\star\pm\pi \equiv z^{\pm}$).   The corresponding energy $\mathcal{A}$ is independent of $z^\star$ to this level of approximation; in practice the collar location becomes sensitive to fine details of the external film distribution \citep{2006-Lister-p311}.

The simulation in Figure~\ref{fig:0} illustrates the redistribution of a film within a rough-walled tube in the absence of external shear.  In this example the initially uniform film evolves into two collars separated by regions in which the film becomes very thin.  The wall roughness is isotropic (Figure~\ref{fig:0}a), with the azimuthally averaged roughness having two prominent maxima within the domain (Figure~\ref{fig:0}b) on which collars form (Figure~\ref{fig:0}c).  The film remains continuous, with fluid continually draining from the ultra-thin film into the neighbouring collars.  The pressure is low and uniform within each collar (Figure~\ref{fig:0}d), rising abruptly across the effective contact lines, which are transverse to the tube axis and spaced a distance $2\pi$ apart across each collar.  Unlike the situation in a perfectly smooth tube, for which collars can migrate axially over long time intervals \citep{2006-Lister-p311}, here collars are pinned by the underlying topography.  Despite significant azimuthal variation in wall roughness, surface tension drives the liquid towards a more axisymmetric film distribution.  The partitioning of the initial fluid volume between the two collars depends on the initial conditions. 

As we illustrate below, sufficiently large shear displaces the collars, ultimately leading to travelling-wave solutions for which collars migrate repeatedly through the domain (because of the periodic boundary conditions).  We identify the loss of stability of all stationary collars to travelling-wave states as the signature of the condition under which liquid can be permanently displaced from a tube, under suitable outlet conditions.

Below we seek to identify the conditions defining the location of collars (such as those in Figure~\ref{fig:0}) and their stability to imposed shear.  In particular, we wish to describe $\tau_c(\omega)$, the largest shear for which a particular tube realisation $a(\theta,z;\omega)$ can support at least one stationary stable collar.  From this we wish to deduce the \textit{collar {{displacement}} probability} $\mathcal{P}(\tau)$.  For a tube drawn randomly from a sample with specified covariance (\ref{covariance}), this is the probability that $\tau_c$ is below a given shear $\tau$, 
\begin{equation}
\label{eq:dispprob}
Pr(\tau_c(\omega)< \tau)\equiv \mathcal{P}(\tau; \eta, r_1, r_2, L, V_0).
\end{equation}
This quantity indicates the likelihood that, over multiple realisations, $\tau$ is sufficiently large to displace all stationary collars of volume $V_0$ or less that may form in a tube with geometric features characterised by $\eta$, $r_1$, $r_2$ and $L$.  A sufficiently large value of $\tau$, ensuring a value of $\mathcal{P}$ approaching unity, can be used as a criterion to ensure reliable removal of liquid from a rough tube under imposed shear.

We first address the simplified case when the roughness is axisymmetric (Sec.~\ref{sec:axisymmetric}), and then return to the non-axisymmetric case in Sec.~\ref{sec:non-axisymmetric}.  In Sec.~\ref{sec:critshear} we determine $\tau_c(\omega)$ by direct simulation and $\mathcal{P}(\tau)$ by a Monte--Carlo method, supplementing these predictions with asymptotic approximations assuming weak roughness $(\eta \ll 1)$.  These yield explicit predictions showing how $\mathcal{P}(\tau)$ is related to the governing parameters, which we extend to liquid plugs in Sec.~\ref{sec:plugs}.

\section{Axisymmetric roughness}
\label{sec:axisymmetric}

\begin{figure}
\begin{center}
\begin{overpic}[width=0.9\linewidth]{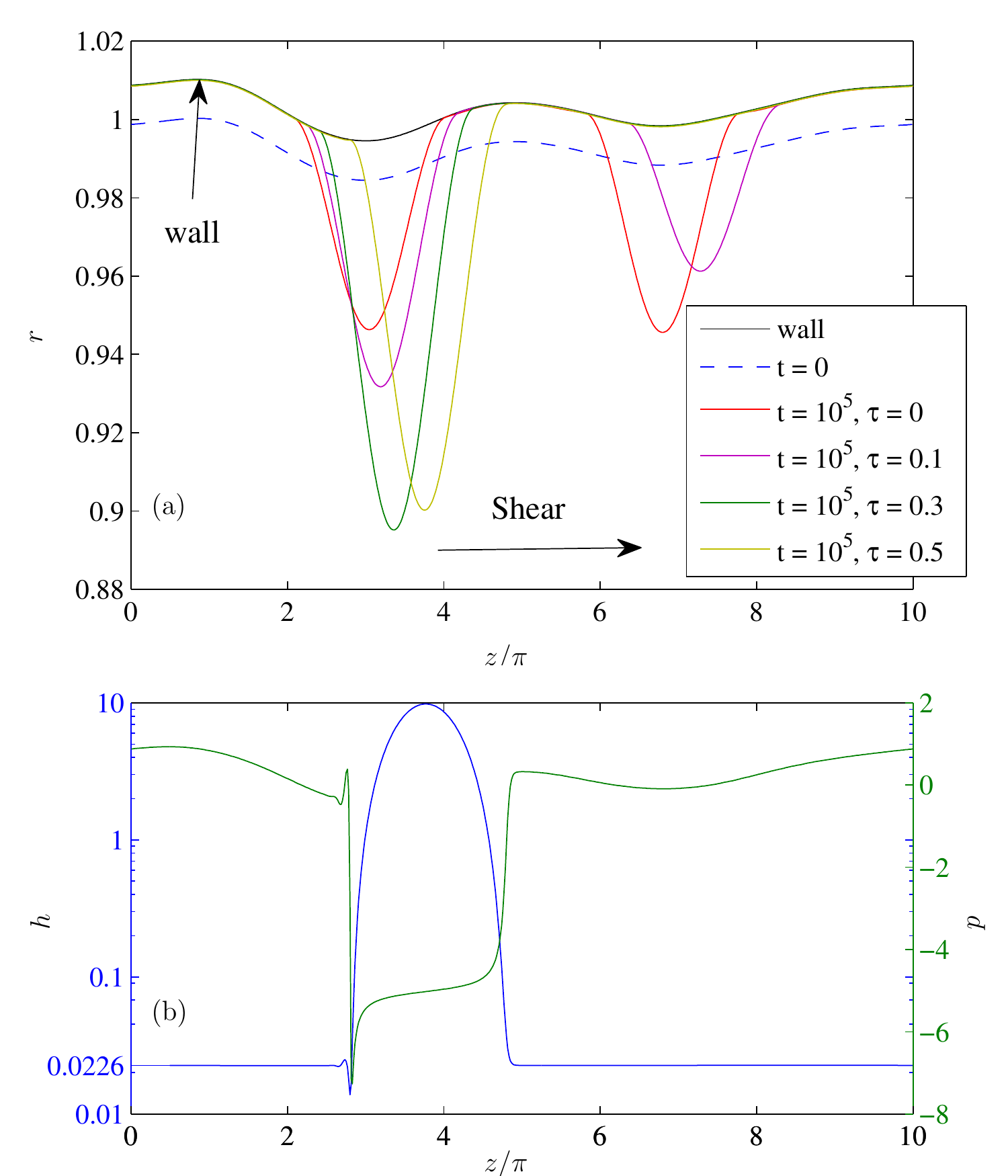}
\end{overpic}
\caption{(a) Solutions of (\ref{EvolutionEqAxis}) with $\eta=1$, $L=10\pi$, $r_2=2\pi$, using a single realisation of wall shape, for $\tau=0$, 0.1, 0.3,  0.5.  The system evolves from an initially uniform film (dashed) to form one or more quasi-steady collars at $t=10^5$.  The gas-liquid interface $r=1-\epsilon g$ and wall $r=1-\epsilon \eta a$ are plotted using radial locations from the tube centreline, taking $\epsilon=0.01$.  (b) The film thickness $h$ (blue, on a logarithmic scale) and pressure $p$ (green, on a linear scale) at $t=10^5$, for $\tau=0.5$. }
\label{ShearCollar}
\end{center}
\end{figure}

We first consider axisymmetric roughness, for which $a(\theta, z) = a(z)$. Equation~(\ref{EvolutionEq}) reduces to
\begin{align}\label{EvolutionEqAxis}
h_t + \left(\tfrac{1}{2} \tau h^2 - \tfrac{1}{3}h^3 p_z\right)_z = 0, \quad p = -(g +g_{zz}), \quad g = h+ \eta a(z).
\end{align}
In the absence of shear ($\tau = 0$), numerical simulations on a particular wall realisation show the film evolves to a quasi-steady state (Figure~\ref{ShearCollar}a), with collars sitting symmetrically on localised constrictions of the tube.   The collars are stationary, with fluid draining into them very slowly from the neighbouring ultra-thin films.  A small increase in shear displaces the collars slightly to the right (Figure~\ref{ShearCollar}a), {{\hbox{i.e.} the downstream direction relative to the imposed shear,}} with each {collar} remaining pinned on a constriction.  (In the present example the simulation was always run from the same initial condition in the presence of shear, leading to different partitioning of fluid between the two collars.)  A further increase to $\tau=0.3$ destabilises the collar on the gentler constriction, so that at large times almost all the fluid accumulates in a single collar on the sharper constriction (Figure~\ref{ShearCollar}a).  A further increase in shear displaces the collar further to the right, preserving its length but moving the rear {{(upstream)}} contact line closer to the tip of the constriction (Figure~\ref{ShearCollar}a, $\tau=0.5$).  For the particular wall shape used in these simulations, no steady collar solution was found for $\tau> 0.5045$; instead travelling-wave disturbances swept through the domain (not shown), reflecting the capacity of shear to displace fluid obstructions from the tube.  The structure of the stationary collar for $\tau=0.5$ is illustrated in Figure~\ref{ShearCollar}(b); the film exterior to the collar becomes uniform at sufficiently large times, with a small capillary wave evident close to the {{rear}} (upstream) contact line of the collar.  The pressure field exterior to the collar reflects the curvature of the wall, and falls abruptly across each contact line.  Stirring of the fluid within the collar by the imposed shear is accommodated by a weak pressure gradient within the collar.  We now analyze the stationary structures in more detail.

\subsection{Collar locations under zero shear}
\label{sec:zero}

For $\tau=0$, at large times there is a weak flux through the effective contact lines of the collar, driven by slow draining from the adjacent film.  We address here the equilibrium structure itself, enforcing uniform pressure in the collar and imposing zero effective contact angle at each contact line.  Thus we solve $p=-(1+\partial_z^2)(h+\eta a)$ with $h=h_z=0$ at $z=z^{\pm}$, {where $z^\pm$ denote the contact line locations with $z^-<z^+$}.  The collar solution may be written as $h=-\eta a + A\cos(z-z^\star)-P$ for $z^-<\{z, z^\star\}<z^+$, where $A$, $z^\star$, $z^\pm$ and $P$ are to be determined.  The four contact-line conditions plus the volume constraint $\int_{z^-}^{z^+} h\,\mathrm{d}z=V$ for some $V\leq V_0$ are sufficient to determine the five unknowns and hence the set of possible collar locations on a rough wall.   

It is helpful to simplify this problem by assuming $0<\eta \ll 1$.  For $\eta=0$, $z^\star$ is undetermined, $z^\pm=z^\star\pm \pi$ while $A=-P=V/4\pi^2$, consistent with (\ref{collarsol0}).  A regular expansion of $A$, $P$ and $z^\pm$ in powers of $\eta$, using  
\begin{equation}
h=(V/4\pi^2)(1+\cos(z-z^\star)) + \eta\left[ -a+A_1\cos(z-z^\star)-P_1\right] +O(\eta^2),
\end{equation}
gives from the contact-line conditions $a^\pm +A_1+P_1=0$ and $(V/4\pi^2) z_1^\pm = a_z^\pm$, where $a^\pm\equiv a(z^\star\pm \pi)$.  The volume constraint implies $P_1=-(1/2\pi) \int_{z^\star-\pi}^{z^\star+\pi} a\,\mathrm{d}z$.   Thus $z^\star$ is determined by the condition that the contact lines meet the wall at identical radial distances from the tube centreline, 
\begin{equation}
a(z^\star-\pi)=a(z^\star+\pi).  
\label{eq:collar0}
\end{equation}
For a rough wall of length $L$ this condition identifies a finite number of possible collar locations, a subset of which are realised in practice.  The collar width becomes
\begin{equation}
z^+-z^-=2\pi+\frac{4\pi^2 \eta}{V_0}(a_z^+ -a_z^-),
\end{equation}
implying for example that a collar that sits on a constriction, for which $a_z^->0$, $a_z^+<0$, is shortened, whereas a collar that sits in a dilation is elongated.  An energetic argument involving the interfacial energy (\ref{eq:intenergy}) shows how stable collars cannot be greater than $2\pi$ in length (Appendix~\ref{app:energy}).  This is supported by numerical evidence (such as Figure~\ref{ShearCollar}a) that collars sitting on constrictions are stable under zero shear; we found no evidence of stable collars sitting in dilations under zero shear.

The stability of a collar under zero shear relies on the interfacial (capillary) energy (\ref{eq:intenergy}) being at a local minimum.  This is distinct from the stability of such collars to weak forcing by shear, which we now consider.

\subsection{Stationary collars under shear}

To interpret numerical observations in Figure~\ref{ShearCollar} for $\tau>0$ we can again use a perturbation analysis assuming both $\eta$ and $\tau$ are small, using the single-collar solution \labelcref{collarsol0} as a starting point.  As suggested by Figure~\ref{ShearCollar}(b), we anticipate that the collar is connected to an external thin film through thin transition regions at each contact line.  Within each transition region, we expect the dominant balance of terms in (\ref{EvolutionEqAxis}) to be 
\begin{equation}
q=\tfrac{1}{2}\tau h^2-\tfrac{1}{3}h^3 h_{zzz}, 
\label{eq:llode}
\end{equation}
where $q$ is the local volume flux.  Solutions of this inner problem must match the collar (for which $h_{zz}=O(1)$) to the external film (of thickness $h_{\mathrm{shear}}$, say).  Balancing terms suggests $q\sim \tau^5$, $h_{\mathrm{shear}}\sim \tau^2$ and the lengthscale of the transition region is $z\sim \tau$ (where `$\sim$' denotes `scales like').  Meanwhile  weak shear appears to act as a regular perturbation to the collar shape on a substrate of amplitude $\eta$ (Figure~\ref{ShearCollar}a), suggesting that $\tau$ and $\eta$ can be assumed of equivalent magnitude at the point at which collars lose stability. 

As is typical of problems with such transition regions, the fluid thickness swept out of the collar at its downstream contact line ($h_{\mathrm{shear}}$, say) is determined in terms of $\tau$ and the collar volume (and pressure), whereas the transition region at the rear contact line can accommodate fluid of variable thickness as it is swept into the collar.   External to the collar, the ultrathin film satisfies $h_t+\tau h h_z=0$ to leading order, which can be solved directly using characteristics.  Over long times, details of the initial transients are swept downstream into the collar's rear contact line (passing through the periodic domain boundary), while the fluid emerging from the collar's downstream contact line provides a source of fixed thickness; thus, ultimately, $h=h_{\mathrm{shear}}$ uniformly across the tube wall exterior to the collar, as in Figure~\ref{ShearCollar}(b).   We can then seek a uniformly steady solution of (\ref{EvolutionEqAxis}).

We therefore develop an expansion in $\tau\ll 1$, formally assuming $\tau$ and $\eta$ are of comparable magnitude as $\tau\rightarrow 0$.  Details of asymptotic matching are provided in Appendix~\ref{app:shearcollar}; here we summarise the conditions determining the collar shape and location.  Since the ODE in each transition region (\ref{eq:llode}) gives rise to logarithmic terms in the far-field, we describe the collar in $\vert z-z^\star \vert <\pi$ using the expansion
\begin{align}\label{eq:expansion}
h = h_0 + (\tau \log \tau)  h_1 + \tau h_2 + o(\tau), \quad z^{\star} = z^{\star}_0 + (\tau\log\tau) z^{\star}_1 + \tau z_2^\star + o(\tau).
\end{align}
Here $h_0$ is given by \labelcref{collarsol0} and $z^{\star}$ denotes the collar mid-point at which
\begin{equation}
 h_z(z^\star)=0.
\label{eq:max}
\end{equation}
Within the collar, the dominant balance of terms in (\ref{EvolutionEqAxis}) becomes
\begin{align}\label{dobalance}
\tfrac{1}{2} \tau=  \tfrac{1}{3}h p_z,\quad p=-(1+\partial_z^2)(h+ \eta a). 
\end{align}
The flux through the collar is neglected in this approximation but shear induces a pressure gradient associated with an internal recirculating flow, visible in Figure~\ref{ShearCollar}(b), which has an impact on the collar shape.  Since the film thickness in the {transition} and external regions is $O(\tau^2)$, we impose the contact-line conditions $h=0$ up to $O(\tau)$ and $h_z=0$ up to $O(1)$ at $z=z^{\pm}$ (this can be justified \textit{a posteriori} via matching, as explained in Appendix~\ref{app:shearcollar}), together with the volume constraint $V = {2\pi}\int_{z^-}^{z^+}h \, \mathrm{d}z$ for some $V\leq V_0$.  The pressure is expanded following (\ref{eq:expansion}).

The leading-order shape satisfies $p_{0_z}=0$, so that $h_0$ is given by 
\begin{equation}
h_0=A_0(1+\cos (z-z_0^\star)),
\label{eq:collar0A}
\end{equation}
equivalent to (\ref{collarsol0}) with $A_0=V/4\pi^2$.  Likewise, at the following order in the expansion (\ref{eq:expansion}), $p_{1z}=0$.  The requirement that this perturbation has zero net volume and vanishing thickness at each contact line gives $h_1=B_1\sin (z-z_0^\star)$ for some constant $B_1$.  Taylor-expanding the maximum condition (\ref{eq:max}) implies $B_1=z_1^\star A_0$.  This solution describes a lateral translation of (\ref{eq:collar0A}); $z_1^\star$ is determined from matching to the transition regions (for details see (\ref{eq:matchcon})).

At the following order, $p_{2z}={3}/(2h_0)$, describing the stirring flow in the collar generated by external shear.  Integrating, 
\begin{equation}
p_2=P_2 + \frac{3}{2A_0} \tan\left(\tfrac{1}{2}(z-z_0^*)\right),\qquad \vert z-z_0^\star \vert<\pi,
\label{eq:p2}
\end{equation}
{ for some constant $P_2$.}  The pressure becomes singular near each contact line, but this singularity is regularized in each transition region.  Integrating (\ref{dobalance}) to find the collar shape gives 
\begin{subequations}
\label{eq:h2}
\begin{equation}
h_2 = - (\eta/\tau) a(z) + P_2 + A_2 \cos(z-z_0^\star) + B_2 \sin(z-z_0^\star) +\frac{3}{2 A_0}F(z-z_0^\star)
\end{equation}
where
\begin{equation}
\label{eq:fu}
F(u)\equiv u \cos u-\sin u \left(1+ 2 \log \left(\cos \left(\tfrac{1}{2}u\right)\right)\right) \qquad (\vert u\vert\ < \pi),
\end{equation}
\end{subequations}
where $A_2$ and $B_2$ are constants of integration and $F$ is an odd function satisfying $F+F''=-\tan
(\tfrac{1}{2}u)$, $F'(0)=0$ and $F\rightarrow\mp \pi$ as $u\rightarrow \pm \pi$.  Imposing $h_2=0$ as $z-z_0^\star \rightarrow \pm \pi$ (a matching condition justified in Appendix~\ref{app:shearcollar}) determines the leading-order collar position, namely 
\begin{align}\label{shear_collar_position}
\eta \left(a(z_0^\star-\pi) - a(z_0^\star +\pi)\right) = {12\pi^3\tau}/{V}.
\end{align}
This condition reduces to (\ref{eq:collar0}) when $\tau=0$.  It shows how, under weak shear, the quasi-steady collar position is determined by an interplay between wall shape $\eta a(z)$, shear $\tau$ and the collar volume $V$.  Appendix~\ref{app:shearcollar} demonstrates how this outer approximation can be systematically matched to {transition} regions near each contact line, giving the displacement $z_1^\star$, and determining the thickness of the ultrathin film $h_{\mathrm{shear}}$ external to the collar.  

We use (\ref{shear_collar_position}) to consider the stability of collars in the presence of shear.  For a given wall shape over a domain of length $L$, we anticipate a set of discrete solutions, giving $z_0^\star$ in terms of the parameter $\Xi\equiv \tau/(V\eta)$.  Displacement of the collar due to changes in this parameter are given to leading order by 
\begin{align}\label{shear_collar_position1}
\left(a_z^- - a_z^+ \right) z_{0,\Xi}^\star= 12\pi^3.
\end{align}
For $a_z^--a_z^+>0$ $(<0)$, as is the case for a collar sitting on a constriction (in a dilation), the collar is displaced downstream (upstream) by increasing shear, which we conjecture implies stability (instability) under shear.  This conjecture is consistent with observations in Figure~\ref{ShearCollar}.   The threshold condition at which a collar loses stability, which corresponds to a saddle-node bifurcation when collar locations are mapped against $\Xi$ (as illustrated below), is therefore (\ref{shear_collar_position}) subject to $a_z^- = a_z^+$. In other words, the limiting collar locations are defined by points on the wall, an axial distance $2\pi$ apart, for which the radial displacement between front and rear contact lines is locally maximal.  The capacity of a tube in realisation $\omega$ to trap a collar of a given volume is given by the global maximum $a_c^+(\omega)$ among this set, where
\begin{equation}
a_c^\pm(\omega)= \max_{0\leq z_0^\star \leq L} \left [ a(z_0^\star\mp\pi;\omega) - a(z_0^\star\pm\pi;\omega) \right]
\label{eq:ac}
\end{equation}
(treating $a$ as an $L$-periodic function).  Likewise if the direction of shear is reversed, the capacity of the tube to trap a collar is determined by $a_c^-(\omega)$.

\subsection{Critical conditions}

Equation~(\ref{shear_collar_position}) indicates how the shear necessary to support a collar scales with collar volume, supporting the observation (\hbox{e.g.} Figure~\ref{ShearCollar}a) that large collars are more resistant to displacement under shear.  We therefore assume that the collar contains almost all the fluid in the domain (\hbox{i.e.} $V=V_0$), in order to estimate the largest shear at which a stationary collar can exist in a tube in a given tube realisation $\omega$:
\begin{equation}
\tau_c(\omega)=\frac{V_0 \eta a_c^+(\omega)}{12\pi^3}.
\label{eq:tauc}
\end{equation}
This prediction can be tested against the simulation data in Figure~\ref{ShearCollar}, for which $\eta=1$: the largest shear was found {numerically to be} $\tau_c\approx 0.5045$.  The small-$\eta$ asymptotic prediction (\ref{eq:tauc}) gives the critical shear as $\tau_c\approx 0.515$. 

\begin{figure}
\begin{center}
\begin{overpic}[width=0.9\linewidth]{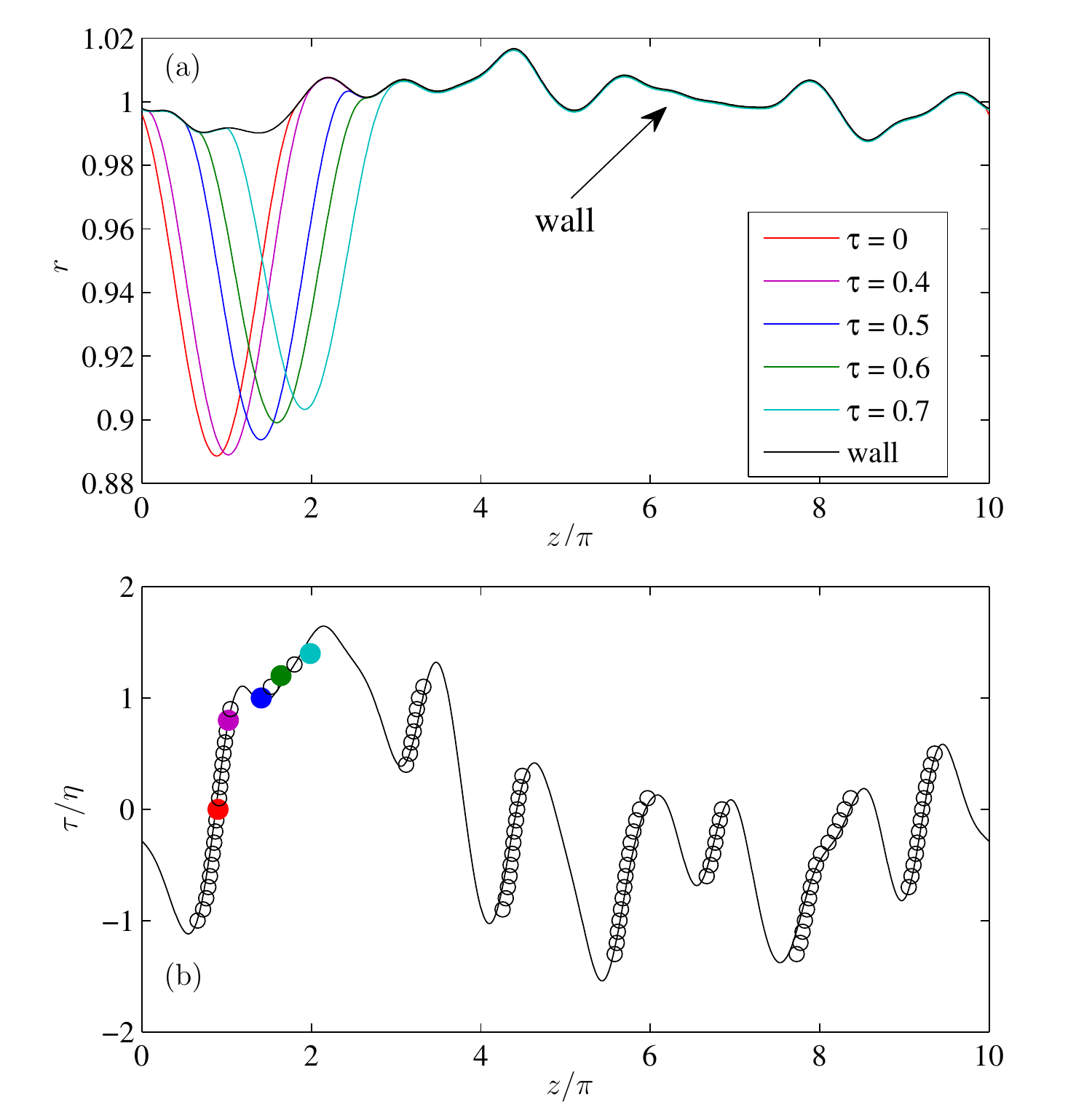}
\end{overpic}
\caption{(a) Stable collars of volume $V=V_0=2\pi L$ on a realisation of a rough wall $a(z; \omega)$ for shear values $\tau=0$, 0.4, 0.5, 0.6, 0.7 and parameters $L=10\pi$, $\eta=0.5$, $r_2 = 1$.  (b) The solid line shows $(V/12\pi^3)[a(z-\pi;\omega)-a(z+\pi;\omega)]$ versus $z/\pi$.    Open symbols identify maxima of stationary collars determined by solving (\ref{EvolutionEqAxis}) for $\tau$ values shown, for forward and reverse shear.  Closed symbols correspond to collars shown in (a).  The alignment of symbols with the line confirms the algebraic condition (\ref{shear_collar_position}).}
\label{fig:collarposition}
\end{center}
\end{figure}

We test these predictions further in Figure~\ref{fig:collarposition}, using a long section of tube with an irregular shape.  For many different values of $\tau$, we solved (\ref{EvolutionEqAxis}) numerically and identified locations at which isolated collars of volume close to $V_0$ could form stationary equilibria.  Different locations were achieved by choosing appropriate initial conditions.   Results of simulations show good agreement with the prediction (\ref{shear_collar_position}) {{(with $V=V_0$)}}, despite the fact that $\eta=0.5$ might be expected to sit outside the range of validity of the small-roughness prediction.  The continuous curve in Figure~\ref{fig:collarposition}(b) arises from evaluating the difference in radial wall location at points an axial distance $2\pi$ apart.  The {segments of this `finite-difference' curve having positive slope} (for which $a_z^->a_z^+$) represent stable collar locations both for $\tau>0$ and when the flow is reversed ($\tau<0$).  The solutions of the algebraic system (\ref{shear_collar_position}) therefore have multiple branches in general; turning points can be identified as saddle-node bifurcation points.  In particular, the largest shear at which a collar is recorded coincides with the predicted global maximum given by (\ref{eq:ac}, \ref{eq:tauc}).  It is evident from the figure that under quasi-steadily reversing shear, a collar may move smoothly back and forth along a single solution branch if the shear is of small amplitude, jump between branches at larger amplitudes or be swept completely out of the tube at sufficiently large amplitude.

We address the distribution of $\tau_c(\omega)$ over many realisations $\omega$ in Sec.~\ref{sec:critshear} below, after consideration of non-axisymmetric wall roughness.

\section{Weak non-axisymmetric roughness}
\label{sec:non-axisymmetric}

To investigate the effect of non-axisymmetric roughness $a(\theta, z)$ on the spatial distribution of annular collars, we have to solve (\ref{EvolutionEq}) numerically, as in Figure~2.  However for small roughness and weak shear a perturbation analysis is revealing.  We seek steady collar solutions by expanding (\ref{EvolutionEq}) following (\ref{eq:expansion}), using 
\begin{subequations}
\begin{align}
h=&h_0(z)+\tau\log \tau h_1(z)+\tau h_2(z,\theta)+o(\tau), \\
p=&p_0(z)+\tau\log \tau p_1(z)+\tau p_2(z,\theta)+o(\tau).
\end{align}
\end{subequations}
We assume $\tau$ and $\eta$ are of comparable magnitude as $\tau\rightarrow 0$, so that the $O(\eta)$ effects of azimuthal wall roughness appear at $O(\tau)$ in the expansion.  The leading-order collar solutions are identical to the axisymmetric case (see (\ref{eq:collar0A})), with $p_0$ and $p_1$ uniform across the collar.  At the following order,
\begin{equation}
-\tfrac{1}{3} h_0^3 p_{2\theta\theta}+ \left(\tfrac{1}{2}h_0^2 - \tfrac{1}{3}h_0^3 p_{2z}\right)_z=0.
\end{equation}
Averaging {azimuthally} using (\ref{eq:axiav}), integrating with respect to $z$ and imposing zero axial flux at this order gives 
\begin{equation}
h_0 \overline{p_{2}}_z=\tfrac{3}{2} \quad\mathrm{where}\quad \overline{p_2}=-(1+\partial_z^2)\left(\overline{h_2}+(\eta/\tau) \overline{a}\right).
\end{equation}
Thus $\overline{p_2}$ is exactly (\ref{eq:p2}) and $\overline{h_2}$ is given by (\ref{eq:h2}) with $a$ replaced by $\overline{a}$.  The contact-line condition  $h_2(\theta,z_0^{\pm}) = 0$ implies $\overline{h_2}(z_0^{\pm}) = 0$.  Therefore, the collar location in a non-axisymmetric rough-walled tube must satisfy 
\begin{align}
\label{maximum}
\eta \left(\overline{a}(z_0^\star-\pi) - \overline{a}(z_0^\star+ \pi) \right) = {12\pi^3\tau}/{V},
\end{align}
which generalises \cref{shear_collar_position} to provide a necessary but not sufficient condition for the existence of a stable collar.  Likewise we modify (\ref{eq:ac}) and (\ref{eq:tauc}) to define the criticality condition in terms of the azimuthally-averaged tube radius
\begin{equation}
\tau_c^\pm(\omega)=\frac{V_0 \eta \overline{a}_c^\pm(\omega)}{12\pi^3},\quad
\overline{a}_c^\pm (\omega)= \max_{0\leq z_0^\star \leq L} \left [ \overline{a}(z_0^\star\mp \pi;\omega) - \overline{a}(z_0^\star\pm \pi;\omega) \right].
\label{shear_collar_position_nonaxis}
\end{equation}
Thus, just as surface tension creates almost axisymmetric structures in a non-axisymmetric tube (Figure~2), we are able to exploit results for an axisymmetric tube to predict outcomes for tubes with two-dimensional roughness.

The present analysis addresses slender collars of length $2\pi$, described by the linearised Young--Laplace equation (\ref{EvolutionEq}b).  Larger collars, described by the nonlinear Young-Laplace equation, become shorter as they grow fatter \citep{1972-Everett-p125, jensen2000}; accordingly their stability will be determined by the spatial fields $\overline{a}(z_0^\star\mp \ell;\omega)-\overline{a}(z_0^\star\pm \ell; \omega)$ for some $\ell \leq 2\pi$.

{Having generalised the stability criterion to non-axisymmetric tubes, we now investigate the properties of the maximum asperity $\overline{a}_c^\pm$ in (\ref{shear_collar_position_nonaxis}) and evaluate the collar displacement probability (\ref{eq:dispprob}).}

\section{Critical shear over multiple realisations}
\label{sec:critshear}

To determine the mean and variance of the critical shear $\tau_c$, when sampled over many tube realisations, we can use Monte--Carlo simulation, repeatedly solving \labelcref{EvolutionEq}.  We can supplement this time-consuming approach by exploiting the algebraic condition \labelcref{shear_collar_position_nonaxis}, valid for small wall roughness, which provides an upper bound on the shear at which stationary collars can be supported within a tube.  In addition to applying Monte--Carlo sampling directly and cheaply to (\ref{shear_collar_position_nonaxis}), we can also approximate the distribution of $\overline{a}_c(\omega)$ directly as follows. 

The azimuthal averaged random field $\overline{a}(z,\omega)$ is Gaussian with mean zero and covariance (also azimuthally averaged)
\begin{align}\label{covariance_averaged}
\overline{k}(z,z') = K(r_1) \textrm{exp}\left[-\frac{1}{2}\left(\frac{\sin(\pi(z - z')/L)}{\pi r_2/L}\right)^2\right], 
\end{align}
where
\begin{align}\label{Fr1}
K(r_1) &\equiv \frac{1}{4\pi^2} \int_0^{2\pi}\int_0^{2\pi}\textrm{exp}\left[-\frac{1}{2}\left(\frac{\sin((\theta_1 - \theta_2)/2)}{r_1/2}\right)^2\right]\dif\theta_1\dif\theta_2 \nonumber \\
 &=  \frac{1}{2\pi^2} \int_0^{2\pi}\textrm{exp}\left[-\frac{1}{2}\left(\frac{\sin(\theta/2)}{r_1/2}\right)^2\right]\left(2\pi-\theta\right)\dif\theta.
\end{align}
$K_1$ increases monotonically with respect to $r_1$, from $K(r_1)\approx r_1/\sqrt{2\pi}$ as $r_1\rightarrow 0$ (an approximation obtained via steepest descents) towards unity for $r_1\gg 1$.  The {differenced} random field $\overline{a}(z) - \overline{a}(z+\ell)$ (with $\ell=2\pi$ for a slender collar) is also Gaussian, with mean zero and covariance
$\overline{k}_d(z,z') = K(r_1) G(z - z')$, where
\begin{multline}
\label{eq:gg}
G(z) = 2\textrm{exp}\left[-\frac{1}{2}\left(\frac{\sin(\pi z/L)}{\pi r_2/L}\right)^2\right] - \textrm{exp}\left[-\frac{1}{2}\left(\frac{\sin(\pi(z -\ell)/L)}{\pi r_2/L }\right)^2\right]  \\ - \textrm{exp}\left[-\frac{1}{2}\left(\frac{\sin(\pi(z +\ell)/L)}{\pi r_2/L}\right)^2\right].
\end{multline}
Thus the {differenced} field has variance $\sigma_d^2 \equiv K(r_1)G(0)$.  We normalise the field by defining $\overline{b}(z) = (\overline{a}(z) - \overline{a}(z+ \ell))/\sigma_d$, which has covariance $G(z)/G(0)$ depending only on $r_2$ and $L$.  We then define $m = -G_{zz}(0)/G(0)$, so that $1/\sqrt{m}$ estimates the distance over which the {differenced} field is correlated.  For $r_2 \ll L$, for example, 
\begin{equation}
m=M(r_2)\quad \mathrm{where}\quad M(r)\equiv \frac{1-(1-(\ell^2/r^2))e^{-\ell^2/2r^2}}{r^2(1-e^{-\ell^2/2r^2})}.
\end{equation}
In this case $1/\sqrt{m}\approx r_2$ for $r_2\ll \ell$ and $r_2/\sqrt{3}$ for $r\gg \ell$.  Since $G$ is even in $\ell$ in (\ref{eq:gg}), it does not distinguish between $\overline{a}_c^+$ and $\overline{a}_c^-$, so we shall describe each as $\overline{a}_c$, {which we call the maximum trapping asperity.}

We characterise the properties of $\overline{a}_c(\omega) = \sigma_d \max_{0\leq z <L}\overline{b}(z)$ in terms of its mean and variance over multiple realisations
\begin{align}\label{MV}
\mathbb{E}(\overline{a}_c) = \sigma_d\mathbb{E}\left(\max_{0\leq z <L}\overline{b}\right), \quad \text{Var}(\overline{a}_c) = \sigma_d^2 \text{Var}\left(\max_{0\leq z <L}\overline{b}\right).
\end{align}
These are functions of the correlation lengths $r_1$ and $r_2$ and the domain length $L$.  It is convenient to exploit the approximation due to \citet{2000-Hill-vol147}  of the probability density function $P_R$ of $\max_{0\leq z <L}\overline{b}$ as an exponentiated Rayleigh distribution \citep{madi2009}, of the form
\begin{align}\label{pdf}
P_R(x) = N \left(1 - e^{-{x^2}/2}\right)^{N - 1} x e^{-{x^2}/{2}}\equiv \left[(1-e^{-x^2/2})^N\right]_x, \quad 
N=\frac{\sqrt{m}L }{2\pi}, \quad (x>0).
\end{align}
Here $N$ can be interpreted as the effective number of independent local maxima of $\overline{b}$ in a domain of length $L$, with each maximum having a Gaussian distribution and with one exceeding all the others in magnitude.  For $N\gg 1$ (\hbox{i.e.} $r_2 \ll L$), for which terms raised to high powers typically approach either zero or unity, the cumulative distribution function for $P_R$ can be approximated to leading order by $H(x-(2 \log N)^{1/2})$, where $H$ is the Heaviside function (the step sits where $P_R'(x)=0$), implying 
\begin{equation}
\mathbb{E}(\overline{a}_c)=\sigma_d \int_0^\infty xP_R(x)\,\mathrm{d}x
\approx \sigma_d (2 \log N)^{1/2},
\end{equation}
so that using (\ref{covariancecyl})
\begin{equation}
\mathbb{E}(\overline{a}_c)\approx 2 \sqrt{K_1(r_1) \left(1-e^{-\ell^2/2r_2^2}\right) \log (L\sqrt{M(r_2)}/2\pi)}\quad\mathrm{for}\quad L\gg r_2.
\label{eq:eac}
\end{equation}
When $r_1=r_2=r$, (\ref{eq:eac}) gives the corresponding prediction for the isotropic covariance function (\ref{covariancecyl}).

\begin{figure}
\begin{center}
\begin{overpic}[width=0.9\linewidth]{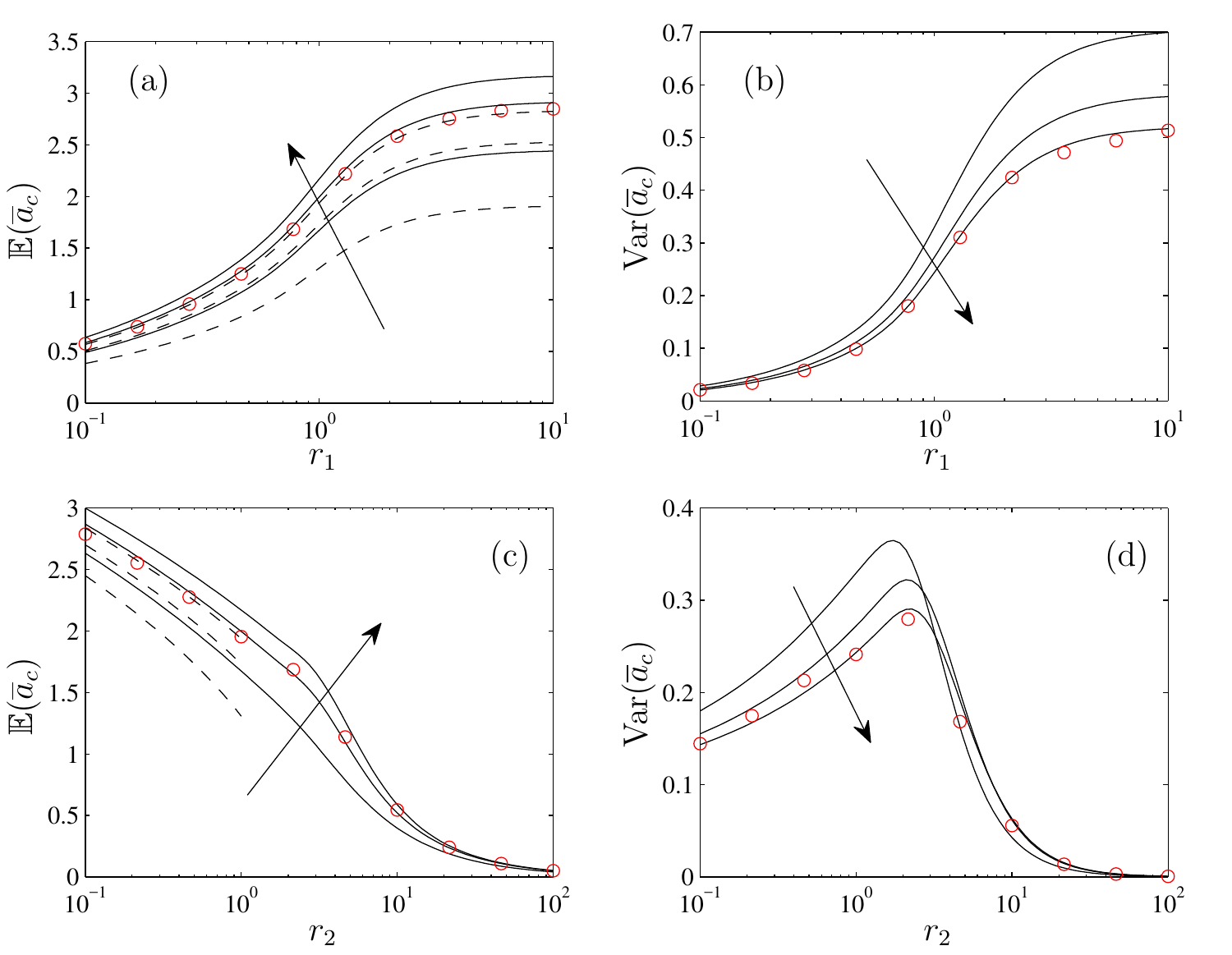}
\end{overpic}
\caption{The mean (left) and variance (right) of the upper bound of the maximum trapping asperity $a_c$ versus the {azimuthal} correlation length $r_1$ with $r_2 = 1$ (top) and {the axial correlation length} $r_2$ with $r_1 = 1$ (bottom).  The solid lines and the dashed lines are calculated from the probability density function \labelcref{pdf}  and the approximated mean \labelcref{eq:eac} with domain length $L = 5\pi$, $10\pi$, $15\pi$ (arrows showing the direction of increasing $L$); the open circles are the Monte--Carlo simulation results directly from \labelcref{shear_collar_position_nonaxis} with $L = 10\pi$.}
\label{MeanVar}
\end{center}
\end{figure}

\Cref{MeanVar} shows how the mean and variance of {the maximum trapping asperity} $\overline{a}_{c}$ depend on the azimuthal correlation length $r_1$, the axial correlation length $r_2$ and the domain length $L$. As $r_1$ increases, both the mean and variance increase to a plateau (Figure~\ref{MeanVar}a, b), showing how a shorter lengthscale of disorder in the azimuthal direction reduces the capacity of the tube to support stationary collars.  The $r_1$-dependence arises through the function $K_1$, given in (\ref{Fr1}).  As the axial correlation length $r_2$ increases, the mean critical shear decreases monotonically (\cref{MeanVar}c) whereas its variance increases and then decreases (\cref{MeanVar}d).  Clearly an increase in $r_2$ will reduce axial gradients of the wall, moving the tube towards the smooth case in which $\overline{a}_c$ tends to zero.  In contrast, very small $r_2$ increases the number of potential pinning sites; here the mean depends on $\sqrt{\log(L/2\pi)+\log(1/r_2)}$, indicating a weak dependence on $L$ as $r_2$ falls.  
Figure~\ref{MeanVar} shows good agreement between the mean estimated from the Monte--Carlo simulations (using $10000$ samples per data point) from \labelcref{shear_collar_position_nonaxis} and the approximate probability density function \labelcref{pdf}, and a reasonable prediction of the variance.  The simplified estimate (\ref{eq:eac}) (assuming $N\gg 1$ in (\ref{pdf})) captures the qualitative form of the mean of $\overline{a}_c$.

\begin{figure}
\begin{center}
\begin{overpic}[width=0.8\linewidth]{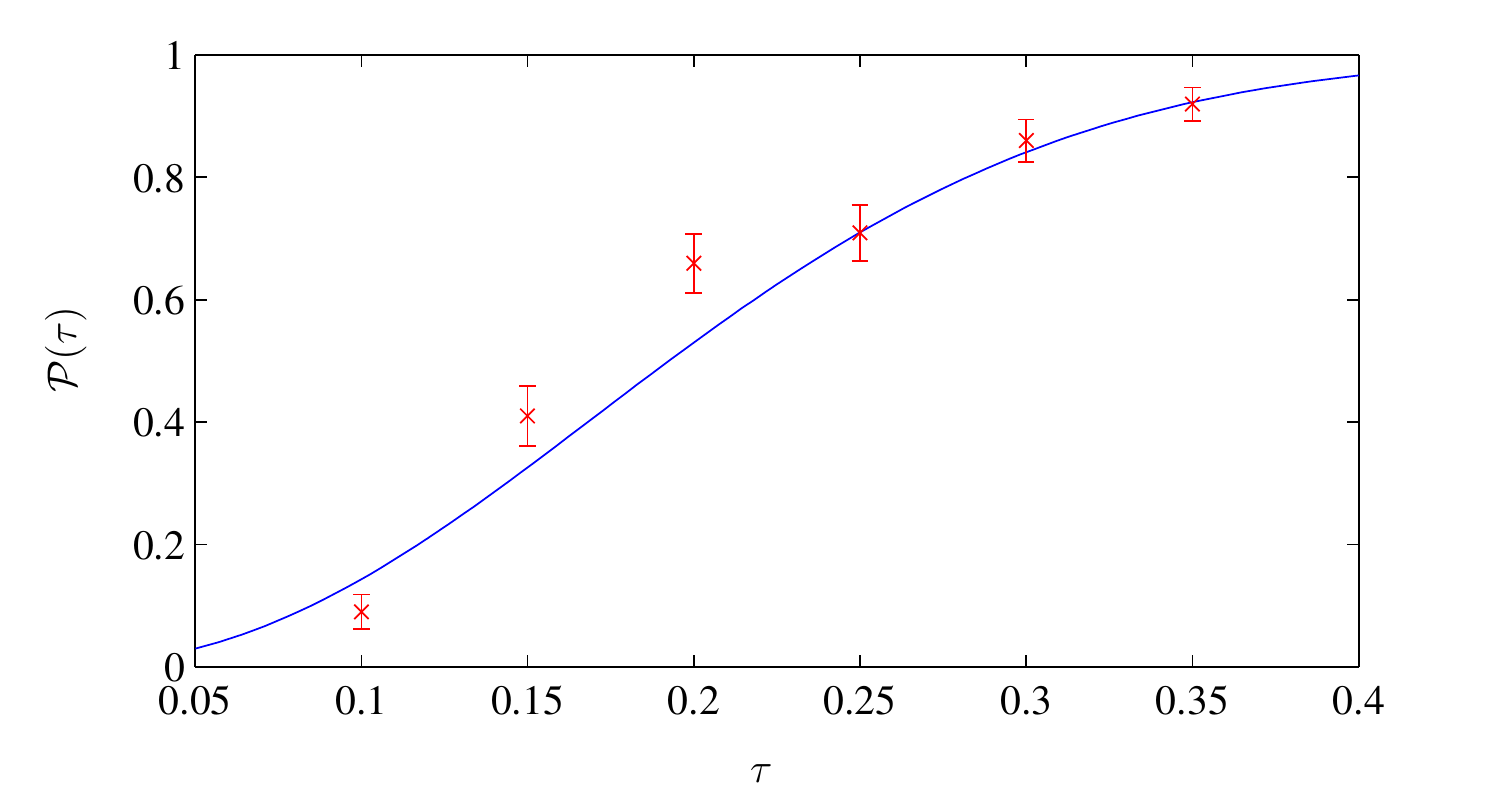}
\end{overpic}
\caption{The probability (frequency) of the critical shear being less than given shear $\tau$, \hbox{i.e.} the collar displacement probability (\ref{eq:dispprob}). The solid line represents \labelcref{PrShear} while the symbols are Monte--Carlo simulations from the axisymmetric \cref{EvolutionEqAxis}. Other parameters are $\eta = 0.5$, $V_0 = 20\pi^2$, $L = 10\pi$ and $r_2 = 5$.   Circles show means and error bars show standard deviations determined using bootstrapping, using 100 simulations at each data point (we resampled with replacement 100 times from these simulations to generate another value of the frequency, and repeating this  procedure $10^4$ times yielded estimates of the mean and the standard deviation).}
\label{PrShearfig}
\end{center}
\end{figure}

Using \labelcref{pdf}, we can approximate the probability (\ref{eq:dispprob}) that the critical shear is less than a given value $\tau$ (corresponding to the probability, over multiple realisations, of the shear $\tau$ being sufficient to displace all quasi-steady collars) as
\begin{align}\label{PrShear}
\mathcal{P}(\tau)& =Pr(\tau_c < \tau) = Pr\left(0<\max_{0\leq z <L}\overline{b}<\frac{12\pi^3\tau}{\eta V_0\sigma_d}\right) 
 = \int_0^{\frac{12\pi^3\tau}{\eta V_0\sigma_d}}P_R(x) \dif x \nonumber \\
& = \left[1-\exp\left(-\frac{72\pi^6\tau^2}{\eta^2 V_0^2\sigma_d^2}\right)\right]^N\,\mathrm{where}\, N=L\frac{ \sqrt{M(r)}}{2\pi},\, \sigma_d=\sqrt{K_1(r) 2(1-e^{-\ell^2/2r^2})}.
\end{align}
Here we assume isotropic covariance (\ref{covariancecyl}) with correlation length $r\ll L$; $\ell =2\pi$ gives the collar length.  \Cref{PrShearfig} shows one example of dependence of the {collar} displacement probability on the given shear $\tau$. Because determining the critical shear for non-axisymmetric wall perturbation is extremely time consuming, we compare \labelcref{PrShear} with that from Monte--Carlo simulations of the axisymmetric evolution equation \cref{EvolutionEqAxis}, finding respectable agreement between them.  The smoothly increasing probability demonstrates the increasing likelihood of eliminating all stationary collars from a tube of a given length as the imposed shear increases; however only in extreme instances can complete removal be guaranteed.  

In a very long tube, for example, for which $N\gg 1$, we expect threshold behaviour with 
\begin{align}
Pr(\tau_c < \tau) & \approx H\left(\frac{12\pi^3\tau}{\eta V_0\sigma_d} -\sqrt{2\log N}\right),
\end{align}
Thus for an isotropic roughness with correlation length $r$, the threshold shear above which collar removal is almost guaranteed takes the form
\begin{align}
\frac{V_0 \eta}{6 \pi^3 (2\pi)^{1/4}} \sqrt{r \log\left(\frac{L}{ 2\pi r}\right)} \quad (r\ll 1 \ll L), \qquad
\frac{V_0 \eta}{6 \pi^2 r} \sqrt{2 \log\left(\frac{L\sqrt{3}}{ 2\pi r}\right)} \quad  (1\ll r \ll L).
\end{align}
The threshold shear diminishes as $r$ becomes very small (because of the increased disorder in the azimuthal direction) and as $r$ becomes very large (because the tube becomes smoother in the axial direction), with an intermediate maximum where the collar length is comparable in magnitude to the correlation length.

\section{Liquid plugs}
\label{sec:plugs}

Finally, and briefly, a similar argument can be developed for liquid plugs.  An isolated plug of radius $a_0^3 V$ in a tube of radius $a_0(1+ \eta a(\theta,z))$ has length $a_0 \ell$ where $\ell=V/\pi+\tfrac{4}{3}$ to leading order in $\epsilon$, where $\ell>2$ (so that the plug has hemispherical ends).  In the presence of a pressure difference $(p^--p^+)\sigma/a_0$, a plug in an axially non-uniform axisymmetric tube can adopt an equilibrium configuration if its hemispherical menisci meet the tube wall with zero contact angle, with the upstream radius being {slightly} shorter than that downstream.  When the wall is non-axisymmetric but weakly rough, we can derive an expression analogous to (\ref{shear_collar_position_nonaxis}), namely (see Appendix~\ref{app:plug})
\begin{align}
\Delta p_c \equiv {(p^- - p^+)_c} = 2\eta \left(\overline{a_1}(z_0^-) - \overline{a_1}\left( z_0^- +\ell  \right)\right)\quad\mathrm{with}\quad  \ell=\tfrac{4}{3} +({V_0}/{\pi}).
\end{align}
We can then exploit directly the estimates of Sec.~\ref{sec:critshear}, such as the plug displacement probability
\begin{equation}
Pr(\Delta p_c<\Delta p)=\left[1-\exp\left( - \frac{(\Delta p)^2}{4 \eta^2 \sigma_d^2 }\right) \right]^N
\label{PrPlug}
\end{equation}
with $N$ and $\sigma_d$ as given in (\ref{PrShear}).  Thus, once again, in a very long tube the displacement probability approaches the step-like form 
\begin{equation}
Pr(\Delta p_c<\Delta p)\approx H \left (\Delta p - 2\eta \sigma_d \sqrt{\log \left(\frac{\sqrt{M} L}{2\pi}\right)}\right).
\label{PrPlugH}
\end{equation}
As before, factors inhibiting plug removal include an increase in roughness amplitude $\eta$, an increase in tube length and a plug length $\ell$ that is of comparable magnitude to the wall correlation length $r$.  

\section{Discussion}

\label{sec:conclusion}

We have investigated the formation and stability of annular collars and plugs in non-axisymmetric rough-walled tubes, focusing our attention primarily on slender collars, for which a comprehensive analysis is possible.  Constrictions in the tube can trap quasi-steady collars, which can then resist displacement under weak shear; likewise, occlusive plugs resist displacement under an imposed pressure gradient.  Trapping of collars is reminiscent of non-wetting drops pinned by contact angle hysteresis \citep{1987-Dussan-vol174, quere2008}, driven drops pinned by surface heterogeneities \citep{savva2013} (and released via sniper bifurcations \citep{thiele2006}), wetting drops spreading on rough surfaces { \citep{2010-savva, espin2015} and capture of fluid from translating droplets by localized asperities \citep{park2017}}.  In the present case, surface tension acting through the film's azimuthal curvature promotes localisation of fluid as a collar which fully wets the wall at its effective contact lines.  A perturbation analysis for weak wall roughness reveals a simple algebraic condition relating collar location to the imposed shear (\ref{maximum}); this appears to be accurate for moderate roughness and is valid for weakly rough non-axisymmetric tubes.  From this we deduced a threshold condition for collar trapping in a particular tube realisation (\ref{shear_collar_position_nonaxis}).

Our primary result is an explicit approximation for the collar displacement probability (\ref{PrShear}) over multiple realisations of a rough-walled tube of finite length, which captures predictions from expensive Monte--Carlo simulations (Figure~\ref{PrShearfig}); it has a straightforward extension for liquid plugs (\ref{PrPlug}) and can be adapted for other scenarios.  This approximation was derived in two steps: first, using asymptotic methods to reduce a model framed as a nonlinear partial differential equation to an algebraic system involving some simple geometric features of the irregular wall shape; second, exploitation of a semi-empirical approximation of the distribution of maxima of a Gaussian random process \citep{2000-Hill-vol147}.  This approach to uncertainty quantification effectively uses a surrogate model, derived using physical principles, to capture the variability of outcomes (rather than rely on purely numerical approaches), a strategy we have successfully adopted in a related problem \citep{xu2016}.  Such low-order descriptions of collar and plug properties are essential in building realistic models of transport in lung airway networks \citep{filoche2015, stewart2015, ryans2016, fujioka2016}, for which stochastic heterogeneity in airway properties is of intrinsic importance.

We focused here primarily on liquid collars, as these are precursors of occlusive liquid plugs in core-annular flows \citep{jensen2000, camassa2014, dietze2015}, although our approach is readily adapted to describe plugs directly (Sec.~\ref{sec:plugs}).  The roughness effects considered here may influence predictions of recent models of liquid plug dynamics in lung airways (\hbox{e.g.} \cite{fujioka2008, ubal2008, magniez2016}) under conditions when plug motion is sufficiently slow for the roughness amplitude to be comparable to the trailing film thickness, although numerous additional factors (surfactant, gravity, etc.) will also need to be accounted for.  For example our work does not address the case in which the roughness has asperities with {axial} lengthscales {shorter than} the film thickness, {for which lubrication theory may not be applicable}.  However our study emphasises that plug or collar dynamics in a real airway (for which geometric properties are known only imperfectly) should sometimes be considered as a stochastic process and clearance of a collar or plug from a real airway can then at best be predicted probabilistically.

\section*{Acknowledgements}
This work was supported by EPSRC grant no. EP/K037145/1.

\begin{appendices}
\crefalias{section}{appsec} 

\section{Interfacial energetics in the absence of shear}
\label{app:energy}

We consider the interfacial energy (\ref{eq:intenergy}) of an isolated collar in an axisymmetric but axially non-uniform tube, in the absence of shear.  Suppose $h$ is positive in $(z_-, z_+)$ and vanishingly small elsewhere.  The stable steady state $h(z)$ minimizes
\begin{equation}\label{local_variational_minimizer}
 \mathcal{A}[h,h_z,z_-,z_+] = \int_{z_-}^{z_+} \left( \tfrac{1}{2}\left(h_z^2 - h^2\right) - \eta \left( a h - a_{z} h_z \right)\right) \text{d} z 
\end{equation}
subject to the volume constraint $\int_{z_-}^{z_+} h\, \text{d}z = V_0$ with  $h(z_-) = h(z_+) = 0$. Introducing a Lagrange multiplier $\lambda$, we construct a new functional
$\mathcal{A}^{*}[h,h_z, z_-, z_+, \lambda] = \mathcal{A}   + \lambda \left(\int_{z_-}^{z_+} h \text{d}z - V_0 \right)  \equiv \int_{z_-}^{z_+} G(z, h, h_z, \lambda) \text{d} z - \lambda V_0$ where $G(z, h, h_z, \lambda)=\tfrac{1}{2} \left(h_z^2 - h^2\right) - \eta \left( a h - a_{z} h_z \right) + \lambda h \equiv G_1(z, \lambda)$.  The variation of the functional $\mathcal{A}^{*}[h(z), z_-, z_+, \lambda]$ up to second order is
\begin{align}
\Delta  \mathcal{A}^{*} &=  \mathcal{A}^{*}[h+\delta h, z_-+\delta z_-, z_+ +\delta z_+, \lambda +\delta \lambda] - \mathcal{A}^{*}[h, z_-, z_+, \lambda] \nonumber \\
&= \int_{z_-+\delta z_-}^{z_++\delta z_+} G(z, h+\delta h, h_z + \delta h_z, \lambda + \delta \lambda) \text{d} z - (\lambda + \delta \lambda) V_0  - \int_{z_-}^{z_+} G(z, h, h_z, \lambda) \text{d} z + \lambda V_0 \nonumber \\
&= \int_{z_-}^{z_+} \left(G(z, h+\delta h, h_z + \delta h_z, \lambda + \delta \lambda) - G(z, h, h_z, \lambda) \right)\text{d} z \nonumber \\
&+ \int_{z_-+\delta z_-}^{z_-} G(z, h+\delta h, h_z + \delta h_z, \lambda + \delta \lambda) \text{d} z + \int_{z_+}^{z_++\delta z_+} G(z, h+\delta h, h_z + \delta h_z, \lambda + \delta \lambda) \text{d} z    - V_0 \delta \lambda.
\end{align} 
The above three integrals can be approximated as
\begin{align}
&\int_{z_-}^{z_+} \left(G(z, h+\delta h, h_z + \delta h_z, \lambda + \delta \lambda) - G(z, h, h_z, \lambda) \right)\text{d} z =  \int_{z_-}^{z_+} \Big(G_h \delta h + G_{h_z} \delta h_z + G_{\lambda} \delta \lambda \nonumber \\
&+ \tfrac{1}{2}\left(G_{hh}(\delta h)^2+G_{h_z h_z}(\delta h_z)^2 + G_{\lambda\lambda}(\delta \lambda)^2 \right) 
+ G_{hh_z} \delta h \delta h_z +G_{h \lambda} \delta h \delta \lambda  + G_{h_z \lambda} \delta h_z \delta \lambda \Big)\text{d} z \nonumber \\
&= \int_{z_-}^{z_+} \Big(G_h  - \frac{\text{d}G_{h_z}} {\text{d} z}\Big)\delta h\, \text{d} z + G_{h_z} \delta h \Big|_{z_-}^{z_+}+ \delta \lambda\int_{z_-}^{z_+} G_{\lambda} \,\text{d} z \nonumber \\
&+ \int_{z_-}^{z_+}\Big( \tfrac{1}{2} (G_{hh}(\delta h)^2+G_{h_z h_z}(\delta h_z)^2 + G_{\lambda\lambda}(\delta \lambda)^2 ) + G_{hh_z} \delta h \delta h_z +G_{h \lambda} \delta h \delta \lambda + G_{h_z \lambda} \delta h_z \delta \lambda\Big)\text{d} z,
\end{align}
\begin{align}
&\int_{z_-+\delta z_-}^{z_-} G(z, h+\delta h, h_z + \delta h_z, \lambda + \delta \lambda) \text{d} z = - \delta z_- G(z, h+\delta h, h_z + \delta h_z, \lambda + \delta \lambda) _{z=z_-+ \theta_1 \delta z_-} \nonumber \\
&\approx - \delta z_- \Big(G+ G_h \delta h + G_{h_z} \delta h_z + G_{\lambda} \delta \lambda \Big)_{z=z_-+ \theta_1 \delta z_-} \nonumber \\
&\approx - \delta z_- \Big(G+  \theta_1 \delta z_-G_{1z}\Big)_{z=z_-}- \delta z_- \Big(G_h \delta h + G_{h_z} \delta h_z + G_{\lambda} \delta \lambda \Big)_{z=z_-} \nonumber \\
&= - \delta z_- G\Big|_{z=z_-}- \delta z_- \Big(\theta_1 \delta z_-G_{1z}+G_h \delta h + G_{h_z} \delta h_z + G_{\lambda} \delta \lambda \Big)_{z=z_-}
\end{align}
and
\begin{align}
&\int_{z_+}^{z_++\delta z_+} G(z, h+\delta h, h_z + \delta h_z, \lambda + \delta \lambda) \text{d} z
&\approx \delta z_+ G\Big|_{z=z_+}+\delta z_+\Big(\theta_2 \delta z_+ G_{1z}+G_h \delta h + G_{h_z} \delta h_z + G_{\lambda} \delta \lambda \Big)_{z=z_+}
\end{align}
where $0< \theta_1, \theta_2 <1$.  The boundary conditions require that $h_{z=z_\pm}=0$, $(h+\delta h)_{z=z_\pm+\delta z_\pm} =0$, which gives approximately
\begin{align}
\label{eq:apbc}
\delta h \Big|_{z=z_\pm} \approx - \Big(\delta z_\pm h_z  + \tfrac{1}{2} (\delta z_\pm)^2h_{zz} + \delta z_\pm \delta  h_z\Big)_{z=z_\pm}.
\end{align}
Thus, the first variation of $\Delta \mathcal{A}^{*}$ is
\begin{align}
\int_{z_-}^{z_+} \Big(G_h  - \frac{\text{d}G_{h_z}}{\text{d} z}\Big)\delta h\,\text{d} z - \Big(G - h_z G_{h_z}\Big)_{z=z_-} \delta z_- + \Big(G - h_z G_{h_z}\Big)_{z=z_+} \delta z_+ +  \Big( \int_{z_-}^{z_+} G_{\lambda} \text{d} z -V_0 \Big)\delta \lambda.
\end{align}
For the first variation to vanish for arbitrary $\delta h$, $\delta z_\pm$ and $\delta \lambda$, we must have
\begin{align}
G_h  - \frac{\text{d}G_{h_z}}{\text{d} z} = 0, \quad \Big(G - h_z G_{h_z}\Big)_{z=z_\pm}=0, \quad  \int_{z_-}^{z_+} G_{\lambda}\, \text{d} z -V_0 = 0,
\end{align}
which recovers (as expected) $\lambda = (1+\partial_z^2)(h+\eta a)$, $h_z(z_\pm)=0$, and $\int_{z_-}^{z_+} h \,\text{d} z =V_0$ with the original boundary conditions $h(z_\pm) = 0$; we may therefore identify $\lambda$ as $-p$.   This problem yields the stationary collar solutions discussed in Sec.~\ref{sec:zero}.

We now address the stability of such solutions by considering the second variation of $\mathcal{A}^{*}$, which is
\begin{align}
& - \Big( \tfrac{1}{2} (\delta z_+)^2h_{zz}G_{h_z} + \delta z_+ \delta  h_z G_{h_z}\Big)_{z=z_+} + \Big( \tfrac{1}{2} (\delta z_-)^2h_{zz}G_{h_z} + \delta z_- \delta  h_z G_{h_z}\Big)_{z=z_-}\nonumber \\
&+ \int_{z_-}^{z_+}\Big(\tfrac{1}{2}(G_{hh}(\delta h)^2+G_{h_z h_z}(\delta h_z)^2 + G_{\lambda\lambda}(\delta \lambda)^2) + G_{hh_z} \delta h \delta h_z +G_{h \lambda} \delta h \delta \lambda + G_{h_z \lambda} \delta h_z \delta \lambda \Big)\text{d} z \nonumber \\
&- \delta z_- \Big(\theta_1 \delta z_-G_{1z}+G_h \delta h + G_{h_z} \delta h_z + G_{\lambda} \delta \lambda \Big)_{z=z_-} +\delta z_+\Big(\theta_2 \delta z_+ G_{1z}+G_h \delta h + G_{h_z} \delta h_z + G_{\lambda} \delta \lambda \Big)_{z=z_+} \nonumber \\
& =  \tfrac{1}{2}\int_{z_-}^{z_+}\Big(-(\delta h)^2+(\delta h_z)^2  +2 \delta h \delta \lambda\Big)\text{d} z- \Big( \tfrac{1}{2} (\delta z_+)^2h_{zz}G_{h_z} \Big)_{z=z_+} + \Big( \tfrac{1}{2} (\delta z_-)^2h_{zz}G_{h_z} \Big)_{z=z_-}\nonumber \\
&- \delta z_- \Big(\theta_1 \delta z_-G_{1z}+G_h \delta h  + G_{\lambda} \delta \lambda \Big)_{z=z_-} +\delta z_+\Big(\theta_2 \delta z_+ G_{1z}+G_h \delta h  + G_{\lambda} \delta \lambda \Big)_{z=z_+} \nonumber \\
& =  \tfrac{1}{2}\int_{z_-}^{z_+}\Big(-(\delta h)^2+(\delta h_z)^2  +2 \delta h \delta \lambda\Big)\text{d} z- \Big( \tfrac{1}{2} (\delta z_+)^2h_{zz}\eta a_{z} \Big)_{z=z_+} + \Big( \tfrac{1}{2} (\delta z_-)^2h_{zz}\eta a_{z} \Big)_{z=z_-}\nonumber \\
&- \delta z_- \Big(\theta_1 \delta z_-h_{zz}\eta a_{z}+(-\eta a + \lambda) \delta h  \Big)_{z=z_-} +\delta z_+\Big(\theta_2 \delta z_+ h_{zz}\eta a_{z}+(-\eta a + \lambda) \delta h  \Big)_{z=z_+} \nonumber \\
& \approx  \tfrac{1}{2}\int_{z_-}^{z_+}\Big(-(\delta h)^2+(\delta h_z)^2  +2 \delta h \delta \lambda\Big)\text{d} z- \Big( \tfrac{1}{2} (\delta z_+)^2h_{zz}\eta a_{z} \Big)_{z=z_+} + \Big( \tfrac{1}{2} (\delta z_-)^2h_{zz}\eta a_{z} \Big)_{z=z_-}\nonumber \\
&- \delta z_- \left(\theta_1 \delta z_-h_{zz}\eta a_{z}+(-\eta a + \lambda)\left( \tfrac{1}{2} (\delta z_-)^2h_{zz} + \delta z_- \delta  h_z\right)  \right)_{z=z_-} \nonumber \\
&+\delta z_+\left(\theta_2 \delta z_+ h_{zz}\eta a_{z}+(-\eta a + \lambda) \left(\tfrac{1}{2} (\delta z_+)^2h_{zz} + \delta z_+ \delta  h_z\right)  \right)_{z=z_+} \nonumber \\
&\approx \tfrac{1}{2}\int_{z_-}^{z_+}\Big(-(\delta h)^2+(\delta h_z)^2  +2 \delta h \delta \lambda\Big)\text{d} z + (\delta z_-)^2  \eta \Big(\tfrac{1}{2} -\theta_1\Big)\Big( a_{z}h_{zz}  \Big)_{z=z_-} + (\delta z_+)^2  \eta \Big(\theta_2- \tfrac{1}{2}\Big)\Big( a_{z}h_{zz}  \Big)_{z=z_+},
\end{align}
where we have exploited (\ref{eq:apbc}) and neglected higher order variation. $\mathcal{A}^{*}$ achieves its local minimum if and only if its second variation is positive for every set of $\delta h$, $\delta z_-$, $\delta z_+$ and $\delta \lambda$ which satisfy $(h+\delta h)_{z=z_\pm+\delta z_\pm} =0$.  If we choose $\delta h = \sin \frac{2\pi(z-z_-)}{z_+-z_-}$, $\delta z_- = 0$, $\delta z_+ = 0$ and $\delta \lambda = 0$, and evaluate the second variation of $\Delta \mathcal{A}^{*}$ as $(2\pi - (z_+ - z_-))(2\pi + (z_+ - z_-))/(2(z_+ - z_-))$, then the collar length $z_+ - z_-$ must be less than $2\pi$ if the collar has local minimum surface area.  Equivalently a collar longer than $2\pi$ can be subject to a perturbation for which the second variation of $\mathcal{A}^*$ is negative, and so such a collar cannot be stable.

\section{Asymptotic structure of a sheared collar}
\label{app:shearcollar}

We develop an asymptotic approximation of a steady collar under shear by starting in the {transition} regions, governed by (\ref{eq:llode}).  We set $z=z_{\pm}+\tau\xi$, $h=\tau^2H(\xi)$, $q=\tau^5 Q$, $p=P(\xi)$, so that to leading order $Q=\tfrac{1}{2}H^2+\tfrac{1}{3}H^3 H_{\xi\xi\xi}$.  We assume $H\rightarrow H_\pm$ and $Q=\tfrac{1}{2}H_{\pm}^2$ for $\xi\rightarrow \pm\infty$, to match with the adjacent films outside the collar.  It is convenient to rescale on these film thicknesses, using $H=H_{\pm}\tilde{H}(\tilde{\xi})$, $\xi=H_\pm^{2/3}\tilde{\xi}$, giving 
\begin{equation}
1=\tilde{H}^2+\tfrac{2}{3}\tilde{H}^3 \tilde{H}_{\tilde{\xi}\tilde{\xi}\tilde{\xi}},\quad \tilde{H}\rightarrow 1\quad \mathrm{for}\quad\tilde{\xi}\rightarrow \pm\infty.
\label{eq:ll}
\end{equation}
As a variant of the well-known Landau--Levich problem, it is easily shown by linearisation about the uniform state that (\ref{eq:ll}) has a unique solution (up to translation) at the downstream ($+$) contact line and a one-parameter family of solutions (up to translation) at the { upstream ($-$) contact line}.

We now consider the solutions of (\ref{eq:ll}) where they match the collar, with $\tilde{H}\gg 1$, satisfying
\begin{equation}
\tilde{H}\approx \tfrac{1}{2}C_{\pm}\tilde{\xi}^2+\frac{3}{C_\pm}\tilde{\xi}\log \vert\tilde{\xi}\vert +\beta_\pm \tilde{\xi} + \frac{9}{2C_\pm^3}(\log\vert \tilde{\xi}\vert)^2+O(\log\vert\xi\vert), \qquad (\xi\rightarrow \mp\infty).
\label{eq:llout}
\end{equation}
The constants $\beta_\pm$ are independent of $C_\pm$ and reflect translation invariance; $C_+$ is determined as a nonlinear eigenvalue (taking the value $C_+\approx 2.386$) while $C_-$ parametrizes the solutions at the rear contact line.  We then express (\ref{eq:llout}) in terms of the original variables, writing $\xi=\xi_0^\pm + Z/\tau$ and noting that 
\begin{align}
\log\vert \tilde{\xi}\vert & =\log\left\vert \frac{\xi_0^\pm + Z/\tau}{H_\pm^{2/3}} \right\vert
=\log\left\vert \frac{Z+\tau \xi_0^\pm}{\tau H_\pm^{2/3}} \right\vert
=\log \vert Z\vert + \log\left \vert 1+\frac{\tau \xi_0}{Z} \right\vert - \log \vert\tau H_\pm^{2/3}\vert \nonumber \\
 & = - \log \tau + \log \vert Z\vert - \log H_\pm^{2/3} + O( \vert {\tau \xi_0}/{Z} \vert ),
\end{align}
so that (\ref{eq:llout}) becomes
\begin{multline}
h\approx \tfrac{1}{2}C_{\pm}Z^2/H_{\pm}^{1/3} - \tau \log\tau \left[ {3ZH_\pm^{1/3}}/{C_\pm}\right]  \\
+\tau\left[  
 \frac{3H_\pm^{1/3}}{C_\pm} Z \log \vert Z\vert  + Z \left(
 C_{\pm}H_\pm^{2/3}\xi_0 + \beta_\pm H_\pm^{1/3} - \frac{3 H\pm^{1/3}}{C_\pm} \log H_\pm^{2/3} \right) + o(1)
 \right] + \dots
\label{eq:outin}
\end{multline}
after discarding all terms that are proportional to $\tau^2$.   This motivates the expansion (\ref{eq:expansion}) in the outer region.  

The first two terms in the outer problem (see (\ref{eq:collar0A})), 
\begin{equation}
h\approx h_0+ (\tau\log \tau) h_1 = A_0(1+\cos(z-z_0^\star))+(\tau \log \tau) z_1^\star A_0 \sin(z-z_0^\star)
\end{equation}
can be expanded near each contact line, using $Z=z-z_0^\star\pm\pi$, and matched to (\ref{eq:outin}) to give 
\begin{equation}
A_0=\frac{C_{\pm}}{H_{\pm}^{1/3}},\quad z_1^\star =\frac{3}{A_0^2}.
\label{eq:matchcon}
\end{equation}
Thus the film thickness swept out of the downstream contact line is determined by the collar volume.  In the example of Figure~\ref{ShearCollar}(b), this prediction gives $h_{\mathrm{shear}}=\tau^2 H_{+}\approx 0.0272$, which is respectably close to the numerically determined value $0.0226$ (bearing in mind that the `small' parameters  in this example are $\tau=0.5$ and $\eta=1$).  Likewise, in a strictly steady state, for which $H_-=H_+$, we may also set $C_-=C_+$ to select the structure of the rear contact line.

The $\tau Z\log \vert Z\vert$ and $\tau Z$ terms in (\ref{eq:outin}) can be matched on to the inner expansion of $F(u)$ in (\ref{eq:fu}) as $u\rightarrow \pm \pi$.  Crucially, as there is no $O(1)$ contribution to the $\tau$ terms in (\ref{eq:outin}), the condition $h_2=0$ can then be applied at each contact line to yield (\ref{shear_collar_position}).

\section{Equilibrium liquid plug in a rough tube}
\label{app:plug}

Consider a cylindrical tube with rough wall $r = a(\theta,z) \equiv 1 + \eta a(\theta, z)$ in terms of cylindrical coordinates $\{r, \theta, z\}$; scale lengths on the mean tube radius $a_0$ and pressure on $\sigma/a_0$.  A liquid plug with volume $V_0$ occupies a domain $\mathcal{V}$ within the tube, and is in equilibrium with gas occupying the remainder of the tube.  We impose gas pressures $p_{\pm}$ downstream ($+$) and upstream ($-$) of the plug; the plug's internal pressure $p_0$ is to be determined.  The liquid wets the wall with zero contact angle.  The plug's free surfaces $S_{\pm}$ have the form $r = u_\pm(\theta,z)$ and touch the wall at $z = z^{\pm}(\theta)$.   Their shape is determined by the Young--Laplace law, contact-line conditions and a volume constraint through
\begin{align}\label{governing}
\nabla \cdot \bm{n}_{\pm} = - \delta p_{\pm}, \quad \bm{\nu} \cdot \bm{n}_{\pm} = 1 \quad \mathrm{and} \quad u_{\pm} = a \quad \mathrm{on}\quad z = z^{\pm}(\theta), \quad \int_{\mathcal{V}} dV = V_0.
\end{align}
Here $\delta p_{\pm} = p_{\pm} - p_0$ are pressure differences between the inside and the outside of the free surface, $\bm{n}_{\pm}$ and $\bm{\nu}$ are the unit normals pointing into the gas phase from the free surfaces and the rigid wall respectively, satisfying
\begin{align}
\bm{n}_{\pm} = {(-1, u_{\pm\theta}/u_{\pm}, u_{\pm z})}/{\Delta_{\pm}}, \quad \bm{\nu} = {(-1, a_{\theta}/a, a_z)}/{A}, 
\end{align}
where $\Delta_{\pm}(\theta, z) = \sqrt{1 + (u_{\pm \theta}^2/u_{\pm}^2) + u_{\pm z}^2}$ and $A(\theta, z) = \sqrt{1 + (a_{\theta}^2/a^2) + a_z^2}$. The mean curvature of the free surfaces is given by
\begin{align}
\nabla \cdot \bm{n}_{\pm} & = -\frac{1}{u_{\pm}\Delta_\pm} + \frac{1}{u_{\pm}}\left(\frac{u_{\pm\theta}}{u_{\pm}\Delta_\pm}\right)_{\theta} + \left(\frac{u_{\pm z}}{\Delta_\pm}\right)_z.
\end{align}

When the wall is uniform, $a = 1$, the free surfaces $u_{\pm}(z)$ are hemispherical, satisfying 
\begin{align}\label{straight_sol}
u_{\pm} = \sqrt{1-(z-z^{\pm})^2} \quad (z^- \leq z \leq z^-+1, \quad z^+ -1 \leq z \leq z^+), \quad {\delta p_{\pm}} = 2.
\end{align}
Thus $p_- = p_+$ and the volume constraint gives the plug length as
\begin{align}\label{straight_length}
z^+ - z^- = ({V_0}/{\pi})+\tfrac{4}{3},
\end{align}
Because $z^+ -z^- \geq 2$, we have $V_0 \geq 2\pi/3$.   However the plug location remains undetermined.

For axisymmetric roughness $a = a(z)$, the free surfaces $u_{\pm}(z)$ satisfy
\begin{align}\label{axis_eq}
-\frac{1}{u_{\pm}\Delta_{\pm}} + \left(\frac{u_{\pm z}}{\Delta_{\pm}}\right)_z = -{\delta p_{\pm}}, \quad {1+u_{\pm z}a_z}={\Delta_{\pm} A}  \quad \mathrm{and} \quad u_{\pm}=a\quad \mathrm{on}\quad z= z^{\pm}.
\end{align}
The boundary condition (\ref{axis_eq}b) simplifies to $u_{\pm z} (z^{\pm}) = a_z (z^{\pm})$.   The problem has the exact solution
\begin{align}\label{axi_sol}
u_{\pm} = \sqrt{R_\pm^2-(z-z_0^{\pm})^2} \quad (z^- \leq z \leq z_0^-+R_-, \quad z_0^+ -R_+ \leq z \leq z^+), \quad {\delta p_{\pm}} = 2/R_\pm.
\end{align}
Applying the boundary conditions gives
\begin{align}
{\delta p_{\pm}}= \frac{2}{a(z^{\pm})\sqrt{1 + a_z(z^{\pm})^2}},
\label{eq:c8}
\end{align}
reflecting tangential contact of the hemispherical interfaces with the wall.  The problem is closed by imposing the volume constraint.  When the wall roughness is small, $\eta \ll 1$, a regular expansion of (\ref{eq:c8}) in $\eta$ gives the condition determining the plug location under a given pressure drop directly as
\begin{align}\label{condition}
{p_- - p_+} = 2\eta \left(a_1(z^+) - a_1\left(z^-\right)\right), \quad  z^+=z^-+({V_0}/{\pi})+\tfrac{4}{3}.
\end{align}
The largest pressure drop at which the tube can trap a stationary plug of volume $V_0$ is then determined by 
\begin{align}
\max_{0<z^-<L} \left(a_1\left(z^-+({V_0}/{\pi})+\tfrac{4}{3}\right) - a_1(z^-)\right).
\end{align}

For non-axisymmetric roughness $a = a(\theta, z)$, the free surfaces $u_{\pm}(\theta, z)$ satisfy
\begin{subequations}
\label{non-axis_eq}
\begin{gather}
-\frac{1}{u_{\pm}\Delta_{\pm}} + \frac{1}{u_{\pm}}\left(\frac{u_{\pm\theta}}{u_{\pm}\Delta_{\pm}}\right)_{\theta} + \left(\frac{u_{\pm z}}{\Delta_{\pm}}\right)_z = -\delta p_{\pm}, \\
{1+u_{\pm \theta}a_{\theta}/u_{\pm}a+u_{\pm z}a_z} = \Delta_{\pm}A \quad \mathrm{and} \quad u_{\pm}=a \quad (z = z^{\pm}(\theta)).
\end{gather}
\end{subequations}
The boundary condition (\ref{non-axis_eq}b) requires $u_{\pm z} = a_z$ and $ u_{\pm \theta}=a_{\theta} $ at $z = z^{\pm}(\theta)$. In the limit of small wall roughness ($a = 1 + \eta a_1(\theta, z)$, $\eta \ll 1$) we construct a regular expansion about the base state (\ref{straight_sol}) of the form
$u_{\pm} = u_{0\pm}(z) + \eta u_{1\pm} (\theta, z) + \dots$, $\delta p_{\pm} = 2 + \eta \delta p_{1\pm} + \dots$, $\Delta=\Delta_{0\pm}+\eta \Delta_{1\pm} + \dots$.   Then $u_1$ satisfies the linear ODE 
\begin{equation}
-\delta p_{1\pm} = \Delta_{0\pm} u_{1\pm} - (z-z_\pm)\frac{{u_{1\pm}}_z}{\Delta_{0\pm}} + \Delta_{0\pm} {u_{1\pm}}_{\theta\theta} + \left( \frac{{u_{1\pm}}_z}{\Delta_{0\pm}^3}\right)_z
\end{equation}
subject to $u_1=a_1$ on $z=z_\pm$.  Taking an azimuthal average, 
\begin{equation}
-\overline{\delta p_{1\pm}} = \Delta_{0\pm} \overline{u_{1\pm}} - (z-z_\pm)\frac{\overline{{u_{1\pm}}}_z}{\Delta_{0\pm}}  + \left( \frac{\overline{u_{1\pm}}_z}{\Delta_{0\pm}^3}\right)_z
\end{equation}
subject to $\overline{u_1}=\overline{a_1}$ on $z=\overline{z_\pm}$.  This is satisfied exactly by $\overline{u_{1\pm}}=B_\pm /u_{0\pm}$, $\overline{\delta p_{1\pm}}=-2B_{\pm}$ with $B_\pm=\overline{a_1}(z_\pm)$.  
Therefore, for weak non-axisymmetric roughness, the leading order position of the equilibrium liquid slug satisfies a stronger form of (\ref{condition}), namely
\begin{align}\label{non-axis_condition}
{p^- - p^+}= 2\eta \left( \overline{a_1}\left(z_0^-+({V_0}/{\pi})+\tfrac{4}{3} \right)-\overline{a_1}(z_0^-)\right).
\end{align}

\end{appendices}

\bibliographystyle{apalike}
\addcontentsline{toc}{section}{\refname} 
\bibliography{References1} 
\end{document}